\documentclass{nature1}

\usepackage{amssymb}
\usepackage{lineno}
\usepackage{amsmath}
\usepackage{gensymb}
\usepackage{graphicx}
\usepackage{xcolor}

\newcommand\aj{ {Astron.J.}}
\newcommand\araa{ {Ann. Rev. Astron. Astrophys.}}
\newcommand\apj{ {Astrophys.J.}}
\newcommand\apjl{ {Astrophys.J.Lett.}}     
\newcommand\apjs{ {Astrophys.J.Supp.}}
\newcommand\aap{ {Astron. Astrophys.}}
\newcommand\mnras{ {Mon. Not. R. Astron. Soc.}}
\newcommand\nat{ {Nature}}

\newcommand{\beq}{\begin{equation}}
\newcommand{\eeq}{\end{equation}}

\newcommand{\Ms}{\textrm{M}_*}
\newcommand{\Msun}{\textrm{M}_\odot}
\newcommand{\kmps}{km~s$^{-1}$}
\newcommand{\MHI}{\rm{M_{H{\textsc i}}}}
\newcommand{\OHI}{{\rm \Omega_{H{\textsc i}}}}
\newcommand{\MB}{{\rm M_B}}

\newcommand{\htwo}{{\rm H_2}}

\newcommand{\hi}{H{\sc i}}
\newcommand{\hii}{H{\sc i}\,21cm}

\title{H{\sc i} 21 cm emission from an Ensemble of Galaxies at an Average Redshift of 1}

\author{Aditya Chowdhury$^{1}$, Nissim Kanekar$^{1}$, Jayaram  Chengalur$^{1}$, Shiv Sethi$^{2}$, K.S. Dwarakanath$^{2}$}

\begin{document}

\maketitle

\begin{affiliations}
 \item National Centre for Radio Astrophysics, Tata Institute of Fundamental Research, \\Pune 411007, India.
 \item Department of Astronomy and Astrophysics, Raman Research Institute, Bangalore, India. 
\end{affiliations}


\begin{abstract}
The baryonic processes in galaxy evolution include gas infall onto galaxies to form neutral atomic hydrogen (H{\sc i}), the conversion of H{\sc i} to the molecular state (H$_2$), and, finally, the conversion of H$_2$ to stars. Understanding galaxy evolution thus requires understanding the evolution of both the stars, and the neutral atomic and molecular gas, the primary fuel for star-formation, in galaxies. For the stars, the cosmic star-formation rate density is known to peak in the redshift range $z \approx 1-3$, and to decline by an order of magnitude over the next $\approx 10$ billion years\cite{Madau14}; the causes of this decline are not known. For the gas, the weakness of the hyperfine H{\sc i} 21cm transition, the main tracer of the H{\sc i} content of galaxies, has meant that it has not hitherto been possible to measure the atomic gas mass of galaxies at redshifts higher than $\approx 0.4$; this is a critical lacuna in our understanding of galaxy evolution. 
Here, we report a measurement of the average H{\sc i} mass of star-forming galaxies at a redshift $z \approx 1$, by stacking\cite{Chengalur01} their individual H{\sc i} 21\,cm emission signals. We obtain an average H{\sc i} mass similar to the average stellar mass of the sample. We also estimate the average star-formation rate of the same galaxies from the 1.4~GHz radio continuum, and find that the H{\sc i} mass can fuel the observed star-formation rates for only $\approx 1-2$ billion years in the absence of fresh gas infall. This suggests that gas accretion onto galaxies at $z < 1$ may have been insufficient to sustain high star-formation rates in star-forming galaxies. This is likely to be the cause of the decline in the cosmic star-formation rate density at redshifts below 1.
\end{abstract}

We report an upgraded Giant Metrewave Radio Telescope [uGMRT\cite{Swarup91,Gupta17}] wide-bandwidth search for \hii\ emission from star-forming galaxies at $z=0.74-1.45$ over a 1.2~square degree region in five sub-fields of the DEEP2 Galaxy Redshift Survey\cite{Newman13}. The low Einstein A-coefficient of the \hii\ transition (relating to spontaneous emission) implies that it is very difficult to detect \hii\ emission from individual galaxies at these redshifts\cite{Fernandez16}. We hence aimed to detect the average \hii\ emission signals from the sample of galaxies, by stacking their \hii\ emission lines\cite{Lah07,Rhee13,Kanekar16,Bera19}. We chose to target the DEEP2 fields due to (1)~the excellent redshift accuracy, corresponding to a velocity uncertainty of $\approx 55$~\kmps, of the DEEP2 Survey\cite{Newman13}, (2)~the large number of galaxies with accurately known spectroscopic redshifts in regions matched to the size of the uGMRT primary beam, and (3)~the DEEP2 redshift coverage of $0.7 \leq z \leq 1.45$, which implies that the \hii\ emission signals from most of the DEEP2 galaxies are redshifted to the frequency range $\approx 580-820$~MHz, and are observable with the uGMRT in a single frequency setting. Our observations cover an interesting epoch in galaxy evolution, overlapping with the peak of star-formation activity ($z \approx 1-3$), and extending to lower redshifts, when the decline in the cosmic SFR density indicates the quenching of star-formation in galaxies.  

We stacked the \hii\ emission from 7,653 blue, star-forming galaxies at $0.74 \leq z \leq 1.45$ within our five uGMRT pointings on the DEEP2 fields, including all blue galaxies whose \hii\ spectra were not affected by systematic effects (see Methods). The \hii\ line stacking was carried out using three-dimensional sub-cubes (two axes of position and a third of velocity) centred on each of the 7,653 galaxies, after smoothing each sub-cube to a spatial resolution of $60$~kpc, and a velocity resolution of $30$~\kmps, and re-sampling each sub-cube onto the same spatial and velocity grid, in the rest-frame of each galaxy. This smoothing and re-sampling was done in order to take into account the cosmological variation of the angular diameter distance with redshift. For each sub-cube, we used the luminosity distance to the galaxy to convert the measured flux density to the corresponding luminosity density. The corresponding pixels (in space and velocity) of the sub-cubes (in luminosity density) of the 7,653 galaxies were then averaged together to produce our final stacked spectral cube. 

Fig.~1 shows the velocity-integrated stacked \hii\ emission signal; the displayed image is the central 270~\kmps\ of the final stacked spectral cube. The stacked \hii\ emission signal is clearly visible in the centre of the image, detected at $\approx 4.5\sigma$ significance. The \hii\ emission is consistent with arising from an unresolved source.

Fig.~2 shows the \hii\ spectrum through the position of maximum flux density of the image of Fig.~1; this too shows a clear detection of the stacked \hii\ emission signal. The \hii\ line luminosity measured from the stacked \hii\ spectrum (see Methods) is ${\rm L_{HI}} = (6.37 \pm 1.42) \times 10^{5}$~Jy~Mpc$^2$~\kmps, at the average galaxy redshift of $\langle z \rangle =1.03$. This implies an average \hi\ mass of $\langle\MHI \rangle = (1.19 \pm 0.26) \times  10^{10} \ \Msun$. (Table 1 provides a summary of our results).

We used simulations to estimate the possible contamination in the above estimate of the average \hi\ mass due to ``source confusion'', i.e. companion galaxies lying within the uGMRT synthesized beam (see Methods). We find that this contamination is negligible, $\lesssim 2$\% for even very conservative assumptions. This is due to the compact uGMRT synthesized beam used for the \hii\ stacking, which has an FWHM of just 60~kpc, similar to the size of an individual galaxy.

The mean stellar mass of the galaxies in our sample is $\langle \Ms \rangle=9.4 \times 10^{9} \ \Msun$ (see Methods\cite{Weiner09}), yielding a ratio of average \hi\ mass to average stellar mass of $\rm \langle \MHI\rangle/\langle\Ms\rangle = 1.26 \pm 0.28$ at $\langle z \rangle=1.03$, i.e. an average \hi\ mass that is comparable to, and possibly larger than, the average stellar mass. This is very different 
from the situation in star-forming galaxies with a similar stellar mass distribution in the local Universe, for which the average \hi\ mass is only $\approx 40$\% of the average stellar mass\cite{Catinella18}. The ratio of \hi\ mass to stellar mass in star-forming galaxies thus appears to evolve from $z \approx 1$ to the present epoch. 

Most star-forming galaxies have been shown to lie on a so-called ``main sequence'' --- a power-law relationship between the SFR and the stellar mass--- at $z \approx 0 - 2.5$, with the amplitude of the power law declining with time\cite{Brinchmann04,Noeske07}. Such main-sequence galaxies form stars in a steady regular manner, and dominate the cosmic SFR density at all redshifts\cite{Rodighiero11}. However, the time for which a galaxy can continue to form stars at its current SFR depends on the availability of neutral gas. This is quantified by the gas depletion time, $\rm t_{dep}$, the ratio of the gas mass (either $\htwo$ or \hi) to the SFR. The $\htwo$ depletion timescale $\rm t_{dep,H_2}$ gives the time for which a galaxy can sustain its present SFR without additional formation of $\htwo$. Conversely, the \hi\ depletion time, $\rm t_{dep,H{\textsc i}} = \MHI/SFR$, gives the timescale on which the \hi\ in a galaxy would be exhausted by star-formation (with an intermediate conversion to $\htwo$). This would result in quenching of the star-formation activity if \hi\ is not replenished in the galaxy, via accretion from the circumgalactic medium (CGM) or minor mergers.

We estimated the average SFR of our 7,653 main-sequence galaxies by stacking their rest-frame 1.4~GHz continuum emission (see Methods), to measure the average rest-frame 1.4~GHz luminosity. We then combined this 1.4~GHz luminosity  with the radio-far-infrared correlation\cite{Yun01} to derive an average SFR of $(7.72 \pm 0.27) \ \Msun/\textrm{yr}$\cite{White07,Bera18}.  Combining this with our average \hi\ mass estimate of $(1.19 \pm 0.26) \times 10^{10} \, \Msun$ yields an average \hi \ depletion time of $\langle{\rm t_{dep,H{\textsc i}}}\rangle = (1.54 \pm 0.35)$~Gyr, for star-forming galaxies at $\langle z \rangle = 1.03$. The \hi\ depletion time is even shorter for the brighter galaxies of the sample, those with absolute B-band magnitude $\rm M_B \leq -21$. To estimate this, we stacked the \hii\ emission from the 3,499 galaxies with $\rm M_B \leq -21$ to obtain an average \hi\ mass of $(1.70 \pm 0.43) \times 10^{10} \, \Msun$, and also stacked their rest-frame 1.4~GHz continuum emission to derive an average SFR of $(16.37 \pm 0.43) \ \Msun/\textrm{yr}$. Combining these measurements, we find that galaxies with $\rm M_B \leq -21$  have $\langle{\rm t_{dep,H{\textsc i}}}\rangle =1.00 \pm 0.25$~Gyr. This is similar to the $\htwo$ depletion time, $\rm t_{dep,H_2} \approx 0.7$~Gyr, obtained for main-sequence galaxies at these redshifts, lying at the upper end of the stellar mass and SFR distributions\cite{Tacconi13}. In the local Universe, star-forming 
galaxies with a similar stellar mass distribution have $\rm t_{dep,H{\textsc i}} \approx 7.8$~Gyr, substantially longer than the $\htwo$ depletion timescale, $\approx 1$~Gyr\cite{Saintonge17}. 
Thus, in the local Universe, main-sequence galaxies can continue to quiescently form stars at the current SFR for $\approx 7.8$~Gyr without the need for fresh gas accretion, as long as there is efficient conversion of \hi\ to $\htwo$ (on timescales shorter than $\rm t_{dep,H_2}$). Conversely, main-sequence galaxies at $z \approx 1$ can sustain their current SFR for only $\approx 1-2$~Gyr, unless their atomic gas reservoir is replenished via gas accretion. This \hi\ depletion time is similar to the timescale on which the cosmic SFR density is observed to decline steeply. This indicates that the quenching of star-formation activity at $z < 1$ is likely to arise due to insufficient gas infall, from the CGM or via minor mergers, resulting in a paucity of neutral gas to fuel further star-formation.

Measuring the redshift evolution of the comoving cosmological \hi\ mass density in galaxies ($\OHI$) is important to understand the global flow of gas into galaxies. In the local Universe, $\OHI$ can be measured from unbiased \hii\ emission surveys\cite{Jones18}, while, at high redshifts, $z \gtrsim 2$, $\OHI$ has been measured from the damped Lyman-$\alpha$ absorbers (DLAs) detected in quasar absorption spectra\cite{Wolfe05}. These studies have shown that $\OHI$ declines by a factor of $\approx 2$ from $z \approx 2.2$ to $z = 0$\cite{Noterdaeme12}. However, the nature of the evolution of $\OHI$ between $z \approx 0.4$ and $z \approx 2.2$ remains unclear, due to the difficulty of carrying out both \hii\ emission studies\cite{Chang10,Kanekar16} and direct DLA surveys\cite{Rao17} at these redshifts.

Our measurement of the average \hi\ mass in a sample of blue, star-forming galaxies allows us to measure $\OHI$ at $z \approx 1$ (see Methods). For this purpose, we use a sub-sample of galaxies with $\MB \leq -20$, for which the DEEP2 survey is expected to be spectroscopically complete at $z \approx 1$\cite{Newman13}. To estimate $\OHI$, we used the known B-band luminosity function for blue galaxies at $z \approx 1$\cite{Willmer06} and the relation between $\MHI$ and $\MB$ from the local Universe [which we find to be consistent with our measurements for galaxies with $\MB \leq -20$; see Methods]. We find that blue, star-forming galaxies with $\MB \leq -20$ contribute $\OHI_{,\rm Bright}=(2.31 \pm 0.58) \times 10^{-4}$ to the total comoving \hi\ mass density at $\langle z \rangle =1.06$. We emphasize that this estimate is a {\it lower limit} to the total $\OHI$ in galaxies at $z \approx 1$, as contributions from \hi\ in faint blue galaxies (and red galaxies) could only increase the total $\OHI$. Extrapolating this estimate to all blue galaxies, again using the B-band luminosity function at $z \approx 1$ and the $\MHI-\MB$ relation of the local Universe, we obtain $\OHI=(4.5~\pm~1.1) \times 10^{-4}$ in blue galaxies at $\langle z \rangle = 1.06$ (see Methods). 

Fig.~3 shows a compilation of $\OHI$ measurements at different redshifts. Our measurement of $\OHI$ at $\langle z \rangle =1.06$ is consistent within the uncertainties with all measurements of $\OHI$ at $z \leq 1$, but is lower (at $\approx 3\sigma$ significance) than the DLA measurement of $\OHI$ at $z \approx 2.15$\cite{Noterdaeme12}. Our results thus indicate that the cosmic \hi\ mass density in galaxies declines substantially by $z \approx 1$, and then remains unchanged at later times. This too indicates that \hi\ in star-forming galaxies is not sufficiently replenished to fuel star-formation at the same level after the peak of star-formation activity

\bibliographystyle{naturemag}

\begin{figure}
\centering
\includegraphics[width=0.8\textwidth]{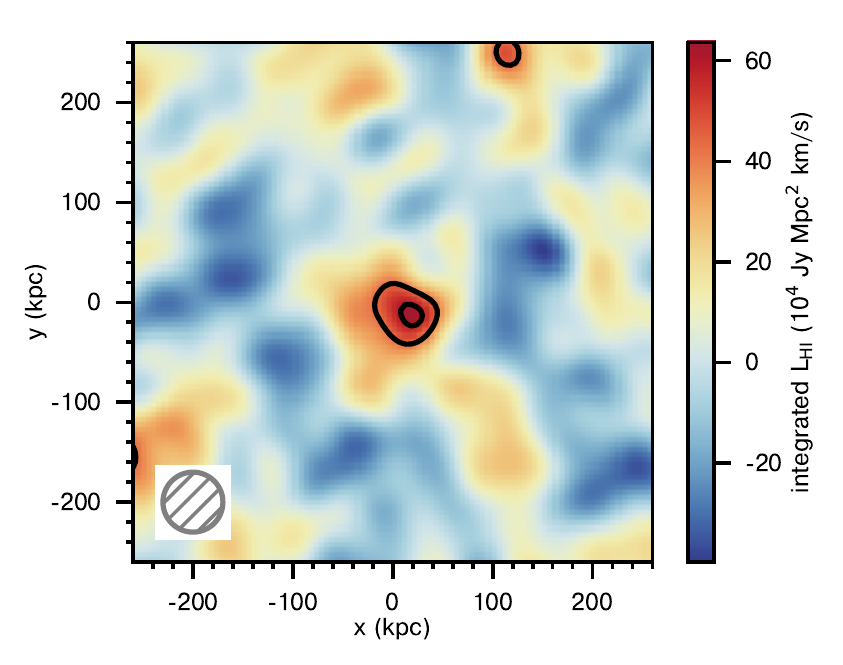}
\caption{\textbf{The final stacked \hii\ emission image.}
This image of the stacked \hii\ line luminosity was obtained by stacking the corresponding spatial and velocity pixels of the sub-cubes centred on each of the 7,653 blue, star-forming galaxies of the sample, covering the velocity range $\pm 135$~\kmps\ around the galaxy redshift (see Methods). The circle in the bottom left indicates the size of the 60~kpc beam (i.e. the spatial resolution). The contour levels are at $3\sigma$ and $4.2\sigma$ statistical significance, where $\sigma$ is the RMS noise on the image. The stacked \hii\ emission signal is clearly detected in the centre of the image, at $\approx 4.5\sigma$ significance, and is statistically consistent with arising from an unresolved source.}
\end{figure}

\begin{figure}
\centering
\includegraphics[width=0.8\textwidth]{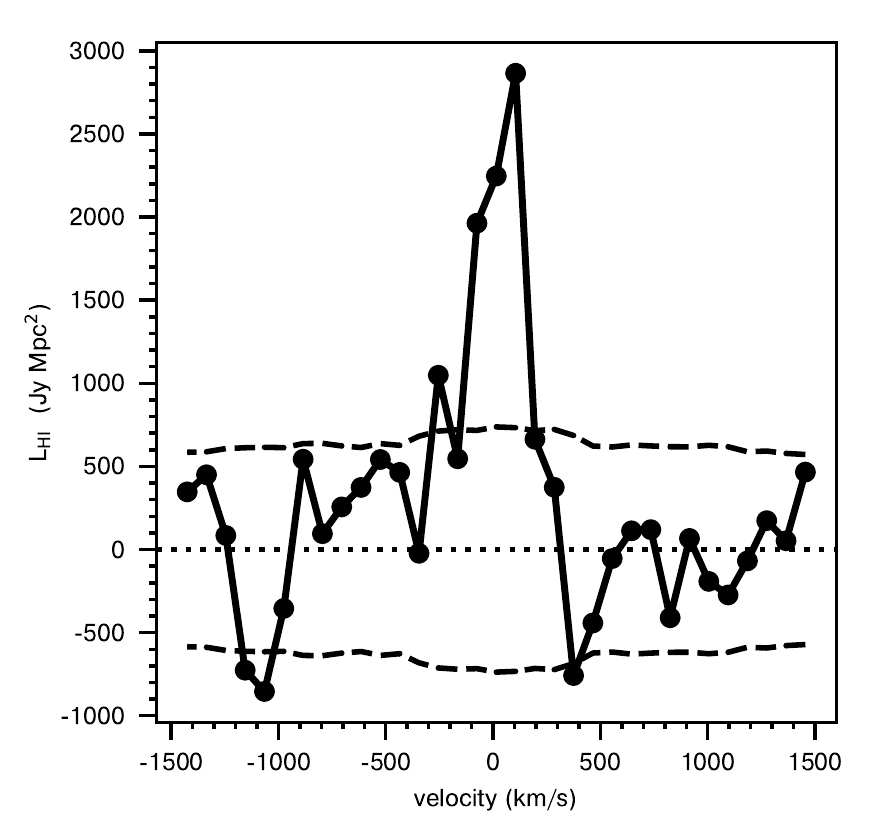}
\caption{\textbf{The final stacked \hii\ spectrum.}
This was obtained via a cut through the location of the peak \hii\ line luminosity in Fig.~1, at a velocity resolution of 90~\kmps. The dashed curve indicates the $1\sigma$ root-mean-square (RMS) noise on the spectrum in each of the 90~\kmps\ velocity channels.  The stacked \hii\ emission signal is clearly detected, at $\approx 4.5\sigma$ significance.}
\end{figure}

\begin{figure}
\centering
\includegraphics[width=0.7\textwidth]{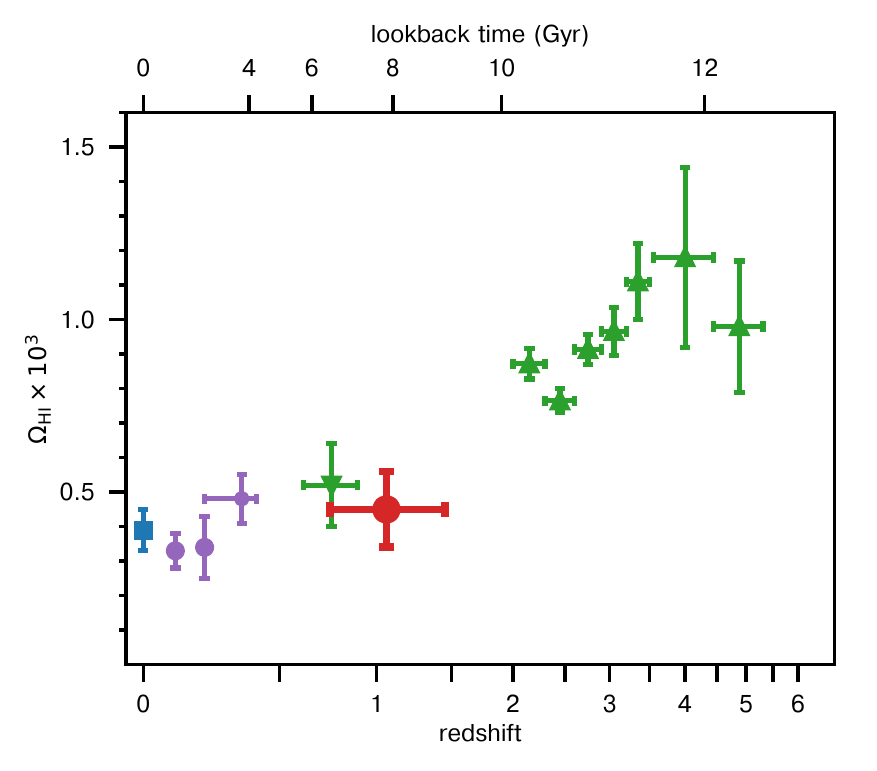}
\caption{\textbf{Redshift (and temporal) evolution of the cosmic \hi\ mass density in galaxies, $\OHI$.}
Vertical error bars indicate $1\sigma$ measurement errors, while horizontal error bars indicate redshift bins. In cases where there are multiple statistically-significant ($\geq 3\sigma$) measurements of $\OHI$ from a single technique at similar redshifts, only the most sensitive result has been included.  The green symbols are measurements from DLAs [triangles]\cite{Noterdaeme12,Crighton15} or Mg{\sc ii} absorbers [inverted triangles]\cite{Rao17}. The blue square is the local Universe measurement of $\OHI$\cite{Jones18}. The purple circles are from low-$z$ \hii\ emission stacking experiments\cite{Rhee13,Bera19}. The filled red circle indicates our estimate of $\OHI$ from all blue galaxies, after correcting for the contribution of faint blue galaxies. Our measurement of $\OHI$ is consistent within the errors with all measurements at $z \leq 1$, but is lower (at $\approx 3\sigma$ significance) than the measurement of $\OHI$ at $z \approx 2.15$ from DLA surveys\cite{Noterdaeme12}.}
\end{figure}

\begin{table}
\centering
\caption{{\bf Details of the sample and the key results of the paper.} The rows are (1)~the number of galaxies whose \hii\ spectra were stacked to detect the average \hii\ emission signal, (2) the redshift range of the stacked galaxies, (3)~their average redshift, $\langle z \rangle$, (4)~their average stellar mass, $\langle \Ms \rangle$, (5)~the average \hi\ mass, $\langle \MHI \rangle$, (6)~the ratio $\langle \MHI \rangle / \langle \Ms \rangle$, (7)~the SFR derived from the rest-frame average 1.4~GHz radio luminosity density, (8)~the \hi\ depletion timescale, $\rm \langle t_{dep,H{\textsc i}} \rangle$, (9)~the comoving cosmological \hi\ mass density in bright galaxies with $\MB \leq -20$ at $\langle z \rangle = 1.06$, $\OHI_{,\rm Bright}$, and (10)~the total $\OHI$ in star-forming galaxies at $\langle z \rangle = 1.06$. } 
\vskip 0.15in
\begin{tabular}{|c|c|}
\hline
    \hline
    Number of Galaxies & $7,653$ \\
\hline
    Redshift range & $0.74-1.45$ \\
\hline
    Mean redshift, $\langle z \rangle$ & $1.03$ \\
\hline
    Mean stellar mass, $\langle \Ms \rangle$ & $9.4 \times 10^9 \ \Msun$\\
\hline
    Mean \hi\ Mass, $\langle\MHI\rangle$ & $(1.19 \pm 0.26)\times10^{10} \ \Msun$\\
\hline
    $\langle\MHI\rangle$/$\langle \Ms \rangle$& $1.26 \pm 0.28$ \\
\hline
    Radio-derived SFR & $7.72 \pm 0.27 \ \Msun/\textrm{yr}$ \\
\hline
    \hi\ depletion timescale, $\langle{\rm t_{dep,H{\textsc i}}}\rangle$ & $1.54 \pm 0.35$ Gyr\\
\hline
   $\OHI_{,\rm Bright}$ at $\langle z \rangle = 1.06$ &  $(2.31~\pm~0.58) \times 10^{-4}$ \\
\hline
   Total $\OHI$ at $\langle z \rangle = 1.06$ &  $(4.5~\pm~1.1) \times 10^{-4}$ \\
     \hline
     \hline
\end{tabular}
\end{table}

\newpage





\begin{methods}
\section{Cosmological Parameters} 
Throughout this work, we use a flat $\Lambda$-cold dark matter ($\Lambda$CDM) cosmology, 
with ($\rm H_0$, $\rm \Omega_{m}$, $\rm \Omega_{\Lambda})=(70$~km~s$^{-1}$~Mpc$^{-1}$, $0.3, 0.7)$.  

\section{The Initial Mass Function and the Magnitude Scale}
The stellar mass and SFR estimates in this work all assume a Chabrier initial mass function (IMF).  
Measurements from the literature that assume a Salpeter IMF have been converted to a Chabrier IMF by 
subtracting 0.2~dex\cite{Madau14b}. All magnitudes are in the AB system. 

\section{Observations and Data Analysis}  
We used the uGMRT Band-4 $550-850$~MHz receivers to observe five sub-fields of the DEEP2 Galaxy 
Redshift Survey\cite{Newman13b} in October--November 2018, with a total observing time of $\approx 90$~hours (see Extended~Data~Table~1).
These five sub-fields are in the DEEP2 fields 3 and 4, at Declination $\approx0\degree$. 
The total on-source time was $\approx 900$~minutes for four of the pointings, and $\approx 450$~minutes 
for the fifth pointing. A bandwidth of 400~MHz was used for the 
observations, sub-divided into 8,192~spectral channels, and centred at 730~MHz. The GMRT 
Wideband Backend was used as the correlator. Observations of one or more of the standard calibrators 
3C48, 3C147 or 3C286 were used to calibrate the flux density scale, while regular observations of 
nearby compact sources were used to calibrate the antenna gains and the antenna bandpass shapes.

The data were analyzed in the Common Astronomy Software Application ({\sc casa}, version~5.4) package\cite{McMullin07}, with the AOFlagger package\cite{Offringa12} additionally used for the detection and excision of radio frequency interference (RFI). The uGMRT has a hybrid antenna configuration with 14 antennas located in a ``central square'', of area $\approx 1$~sq.~km, and the remaining 16 antennas lying along the three arms of ``Y'', providing  baselines out to $\approx$ 25~km. The hybrid configuration provides some insurance against RFI, as RFI decorrelates on the longer baselines. We took advantage of this by entirely excluding the 91 central square baselines from our analysis, working with only the 344 long (i.e. $\gtrsim 1$~km) baselines of the array. 

The antenna-based complex gains and system bandpasses were estimated from the data on the calibrator sources, with our own custom routines developed within the {\sc casa} framework. The algorithms used in these routines are more robust to the presence of RFI in the data, and thus yield a more accurate calibration than the standard {\sc casa} routines. After applying these initial calibrations, the target-source visibilities were smoothed to a spectral resolution of 0.488~MHz for the purpose of continuum imaging; this reduces the data volume by a factor of 10 while avoiding bandwidth smearing of source structures. On each field, we performed multiple iterations of the standard imaging and self-calibration procedure (again using our calibration routines), along with RFI excision, until no further improvement was seen in the continuum image. The imaging at each self-calibration iteration was done using the {\sc tclean} routine, with w-projection\cite{Cornwell08}, and multi-frequency synthesis (2nd-order expansion)\cite{Rau11}. 

At the end of the self-calibration procedure, the fraction of target-source data lost to all time-variable issues within a run (e.g. RFI, temporarily malfunctioning antennas, power failures, etc., but excluding entirely non-working antennas and the 91 central square baselines that were excised at the outset), is $\approx 20-30\%$ for each field. Extended~Data~Figure~1 shows the fraction of data excised due to such time-dependent issues as a function of observing frequency for the entire 4004 minutes of observation; the median fractional data loss across our observing band is $\approx$~20\%.

The final continuum image of each field was created using the {\sc tclean} routine, with Briggs weighting of 0.5, w-projection\cite{Cornwell08}, and multi-frequency synthesis (2nd-order expansion)\cite{Rau11}. A region of radius $0.75$~degrees was imaged for each field, extending far beyond the null of the uGMRT primary beam at our observing frequencies. The RMS noise on our continuum images is $\approx 5-8 \, \mu$Jy/Beam away from bright continuum sources, with synthesized beam (full width at half maximum, FWHM) widths of $\approx 5''$ (see Extended~Data~Table~1). 

We used the {\sc uvsub} routine to subtract all detected radio continuum emission from each 
self-calibrated visibility data set before making the spectral cubes. The cubes were made 
in the barycentric frame (after applying a correction for the shape of the primary beam), 
using natural weighting. Each cube has a channel resolution of $48.83$~kHz, corresponding 
to velocity resolutions of $18-25$~\kmps\ across the frequency band ($820-580$~MHz). The large 
frequency range implies that the FWHMs of the synthesized beam of each cube are different at 
different frequencies, $\approx 3.8'' - 7.5''$, corresponding to a physical size of $30-70$~kpc 
for the redshift range $0.74 - 1.45$.

\section{Sample Selection} 
The DEEP2 Survey used the DEIMOS spectrograph on the Keck~II Telescope to accurately measure the spectroscopic redshifts of 38,000 galaxies at $z \approx 0.70-1.45$, in four regions of the sky\cite{Newman13b}. The redshifts were measured from the O{\sc ii}$\lambda 3727$ doublet, with a high spectral resolution $R=6000$. Both the large number of galaxies and the excellent redshift accuracy 
(corresponding to a velocity uncertainty of $\lesssim 55$~\kmps) of the DEEP2 Survey\cite{Newman13b} are critical to our aim of detecting the stacked \hii\ emission signal. The large number of galaxies increases the signal-to-noise ratio of the stacked \hii\ emission signal, while a redshift accuracy $\lesssim 100$~\kmps\ is important to prevent the stacked signal from being smeared in velocity (e.g.\cite{Maddox13,Elson19}). The DEEP2 Survey targeted galaxies for spectroscopy to a completeness limit of ${\rm R_{AB}}=-24.1$\cite{Newman13b}. This selection criterion favours blue, star-forming galaxies at $z=0.70-1.45$\cite{Willmer06}.

Our sample consists of galaxies in the redshift range $z=0.74-1.45$, for which the rest-frame velocity 
range of $\pm1500$ \kmps\ for the redshifted \hii\ line lies in the frequency range $\approx 580-820$ MHz, i.e. 
in the sensitive part of the uGMRT band. For the five DEEP2 sub-fields, there are 11,370 DEEP2 galaxies 
in the redshift range $0.74 - 1.45$ with reliable redshifts (quality code, Q~$\geq 3$) lying within 
the half-power point of the uGMRT primary beam at the galaxy's redshifted \hii\ line frequency. We 
initially rejected red galaxies, with color ${\rm C}>0$, where ${\rm C}$ is a combination of the 
rest-frame B-band magnitude ${\rm M_B}$ and rest-frame ${\rm U-B}$ color\cite{Willmer06}: 
${\rm C=U-B}+0.032 \times ({\rm M_B}+21.63)-1.014$. C is defined such that the 
C$=0$ line passes through the green valley which separates the blue galaxies from the red ones in the color-magnitude diagram. We note that the R-band selection criterion for the DEEP2 survey preferentially picks out blue galaxies at $z \approx 1$\cite{Willmer06}. As a result, only 1469 galaxies, i.e. $\approx 13$\% of the galaxies of our sample, are red systems, with most of these at the lower end of the redshift coverage. After applying the color selection, there are 9,901 blue galaxies in the sample.

Next, any sample of star-forming galaxies contains contamination from active galactic 
nuclei (AGNs), which form a different population from main-sequence galaxies. The presence
of AGNs in the DEEP2 sample is likely to affect both our SFR and $\langle\MHI\rangle$ estimates. 
Studies of radio sources have found that AGNs typically have rest-frame 1.4~GHz luminosity 
densities ${\rm L_{1.4 GHz}}\gtrsim 2 \times 10^{23}$ W/Hz\cite{Condon02}. We used this 
luminosity threshold to exclude possible AGNs from the DEEP2 sample. This was done by using 
the measured flux density of each DEEP2 galaxy in our continuum images, along with the 
galaxy redshift and an assumed spectral index $\alpha = -0.8$, to estimate its rest-frame 
1.4~GHz luminosity density. All DEEP2 galaxies detected at $\geq 4\sigma$ significance in 
our continuum image, with ${\rm L_{1.4 GHz} \geq 2 \times 10^{23}}$~W/Hz, were excluded 
from the sample. 435 objects were identified as likely AGNs using this criterion, leaving 
us with a sample of 9,466 blue, star-forming galaxies.

\section{\hii\ sub-cubes and spectra}

We extracted three-dimensional \hii\ sub-cubes around the spatial position and redshifted 
\hii\ frequency of each galaxy, covering the rest-frame velocity range $\pm 1500$~\kmps, and 
an angular range of $\pm 38.4$ arcseconds. Each sub-cube was then convolved with a two-dimensional 
Gaussian function to obtain a synthesized beam of angular FWHM corresponding to a physical size of 
$60$~kpc at the galaxy's redshift. In other words, we took into account the relation between 
angular diameter distance and redshift to produce sub-cubes with the same spatial resolution 
($60$~kpc) around each target galaxy. 

The naturally-weighted synthesized beam of each sub-cube deviates substantially from a 
Gaussian beam. This needs to be accounted for while scaling the convolved maps by beam area 
ratios to ensure that these maps are in the correct Jy/beam unit. This correction, 
for each frequency channel in a sub-cube, was applied by (a)~convolving the point spread 
function with the same kernel as was done for the sub-cube, (b)~computing the inverse of 
the value of the central pixel in the convolved point spread function, and (c)~multipying 
the frequency channel of the sub-cube by this factor. The above procedure ensures that the 
central pixel of the convolved point spread function is correctly normalized to unity.

After convolution to the same physical scale with a synthesized beam FWHM of 60~kpc, we regridded 
each sub-cube to a uniform physical pixel size of 5.2~kpc and a spatial range of $\pm 260$~kpc. 
Next, we fitted a second-order spectral baseline to each spatial pixel and subtracted it out. 
This was done to remove the effects of low-level deconvolution errors from continuum sources, as 
well as any residual errors from bandpass calibration. Following this, we interpolated the spectral 
axis in each sub-cube to a single rest-frame velocity grid with a velocity resolution of $30$~\kmps. 

Next, the \hii\ spectrum for each galaxy was obtained by taking a cut through the galaxy's 
location in its sub-cube, covering a velocity range of $\pm 1500$~\kmps\ around the galaxy 
redshift, with a uniform velocity resolution of $30$~\kmps. These spectra were used to 
further screen the galaxy sample that was used for the final stacking analysis, by testing 
each spectrum for non-Gaussian behaviour. This is because the RMS noise of a stacked \hii\ 
spectrum decreases with the number $\rm N$ of individual spectra that have been stacked 
together as $\propto 1/\sqrt{N}$, if the stacked spectra contain no systematic effects or 
correlations. We hence excluded galaxies whose \hii\ spectra are affected by RFI or show 
any signatures of non-Gaussianity, based on the following criteria: 
\begin{itemize}
    \item A spectrum is rejected if more than $15\%$ of its channels have been completely discarded.

    \item A spectrum is rejected if it has a spectral feature of $\geq 5.5\sigma$ significance 
          either at the native velocity resolution ($30$~\kmps), or after smoothing to resolutions of 
         $60$~\kmps\ and $90$~\kmps.

    \item Each spectrum was tested for Gaussianity, using the Anderson-Darling test and the 
          Kolmogorov-Smirnov test,  at the native resolution of $30$~\kmps, and after smoothing to 
          resolutions of $60$~\kmps\ and $90$~\kmps. A spectrum is rejected if it fails either of the 
          tests at any of the three resolutions, with a p-value $<0.0002$. 

    \item Finally, each spectrum was examined for the presence of correlations (e.g. due to a residual 
          spectral baseline) by examining the decrease in the RMS noise after smoothing to coarser velocity 
          resolutions. Specifically, we smoothed each spectrum by a factor of 4 to a resolution of 120~\kmps\ 
	and rejected spectra whose RMS noise decreases by a factor $<1.45$ after the smoothing. 
	This too is effectively a test for non-Gaussianity; the p-value corresponding to our 
	rejection criterion is $\approx 0.0003$.
\end{itemize}

After excising spectra based on the above tests, our sample contains 7,925 galaxies. Finally, 
in order to stack in the image plane at a physical resolution of $60$~kpc, we excluded galaxies 
for which the naturally-weighted synthesized beam at the galaxy's redshifted \hii\ frequency 
corresponds to a physical size $>60$ kpc. This was found to be the case for $272$ galaxies whose 
redshifted \hii\ frequencies lie in spectral channels where data from antennas on the longer 
GMRT baselines have been preferentially excised. After excluding these galaxies, our final galaxy 
sample contains 7,653 blue, star-forming galaxies.

We emphasize that our results do not depend substantially on the thresholds chosen for any of 
the above tests of non-Gaussianity, and also do not change appreciably if we retain the 
galaxies that were rejected due to their large synthesized beam widths.

\section{Stacking the \hii\ emission} 

The stacking of the \hii\ emission was carried out directly on the sub-cubes of individual galaxies, 
rather than merely on the \hii\ emission spectra. This has the advantage that any putative signal 
would be detected both spatially and spectrally, allowing additional tests of its reality (e.g. by 
inspecting off-signal spatial regions for non-Gaussian behaviour). The stacking was carried out by 
first converting the \hii\ line flux density (${\rm f_{HI}}$) of each galaxy to the \hii\ luminosity 
density (${\rm L_{HI}}$) at the galaxy redshift ($z$), using the relation ${\rm L_{HI}}=4\pi \ 
{\rm f_{HI}} \ {\rm D_L}^2/(1+z)$, where ${\rm D_L}$ is the luminosity distance at the redshift of 
the galaxy. We then averaged the luminosity density in the corresponding spatial and spectral pixels 
around each galaxy to obtain the stacked spectral cube in \hii\ luminosity density. Finally, we fitted 
a second-order spectral baseline to each spatial pixel in the stacked cube, after excluding the 
central $\pm 350$~\kmps\ region, and subtracted out this baseline to obtain the final spectral cube. We note that the average \hii\ emission signal is clearly detected even without subtracting the second-order polynomial from the stacked \hii\ spectrum.

We repeated the above procedure with beam FWHMs larger than $60$~kpc and found no evidence for an increase 
in the derived average \hi\ mass. Using a larger beam FWHM increases the RMS noise (and thus decreases 
the signal-to-noise ratio) due to the down-weighting of the longer uGMRT baselines. We hence chose a 
spatial resolution of 60~kpc for our final spectral cube. In passing, we note that 60~kpc is the 
typical diameter of galaxies in the local Universe with an \hi\ mass of $\rm \MHI \approx 10^{10} 
\, M_\odot$\cite{Wang16} (i.e. the average \hi\ mass of our 7,653 galaxies).

The RMS noise on each channel of the stacked \hii\ spectrum was estimated by making 10,000 realizations 
of the stacked spectrum, using bootstrap re-sampling (with replacement) of the 7,653 individual \hii\ 
spectra. We note that the RMS noise in the $\pm 350$~\kmps\ region is slightly higher than the noise in 
other regions. This is due to the fitting of a second-order baseline to the final spectrum excluding this 
velocity range, which has the effect of marginally increasing the noise in the line region. We note that we also estimated the RMS noise on the stacked \hii\ spectrum via a Monte Carlo approach, where the systemic velocity of each galaxy in our sample was shifted by a random offset in the range -1500~\kmps\ to 1500~\kmps\ before their \hii\ spectra were stacked; the error obtained from this approach is consistent with that obtained from bootstrap re-sampling (with replacement).
 
In \hii\ stacking experiments (except for a uGMRT experiment at $z\approx0.34$\cite{Bera19b}), the \hii\ spectra from individual galaxies are typically weighted by the inverse of their variance before stacking them in flux 
density, to optimize the RMS noise on the final stacked spectrum. However, our observations span a large redshift range, $0.74 \leq z \leq 1.45$; stacking in flux density would have the unwanted effect of a bias towards galaxies at the low-redshift end of our coverage. We hence carried out the stacking in \hii\ luminosity density, instead of \hii\ flux density. Further, the RMS noise on the individual \hii\ luminosity density spectra would also be higher for the higher-redshift galaxies, by a factor of $\approx 2$, due to their larger luminosity distances. Hence, the standard approach of weighting by the inverse of the variance would again bias the stacked spectrum towards lower-redshift galaxies. To avoid this, we stacked the \hii\ luminosity density spectra without any weights. Finally, we note that the results obtained on weighting the spectra by the inverse of the variance of the \hii\ luminosity density are consistent (within $1\sigma$ significance) with the results obtained without weighting. Our conclusions thus do not depend on whether we stack with equal weights or use an inverse-variance weighting scheme. 

We also carried out a median stack of the \hii\ emission signals of the 7,653 galaxies in our sample, obtaining an average \hi\ mass estimate of $\MHI = (1.09 \pm  0.29) \times 10^{10} \; \Msun$. This is entirely consistent with our \hi\ mass estimate from the mean stack of $\MHI = (1.19 \pm  0.26) \times 10^{10} \; \Msun$.

\section{Data Quality and Systematic effects} 
We compared the RMS noise on the \hii\ spectrum of each of the 7,653 galaxies of the final sample to the theoretical RMS noise, based on the sensitivity of the uGMRT receivers.  The predicted RMS noise was calculated using the sensitivity curve provided by the uGMRT observatory, after taking into account the average flagged fraction, and the effects of spectral and spatial smoothing. Extended~Data~Figure~2 (blue dots) shows the RMS noise per 30~\kmps\ channel for each galaxy, plotted against the observing frequency, while the red curve shows the theoretical sensitivity curve for a galaxy at the centre of the field (the theoretical sensitivity would be worse for a galaxy away from the field centre, due to the telescope primary beam response). It is clear that the theoretical sensitivity provides a lower envelope to the observed RMS noise values, and that the spread in the RMS noise values is a factor of $\approx 2$ at any given frequency. The spread in the observed RMS noise values is because we have included galaxies out to the half-power point in the telescope primary beam, whose spectra would have an RMS noise two times worse than that of galaxies at the field centre. We thus find that the observed RMS noise values on the individual \hii\ spectra are consistent with the predicted RMS noise. 

Next, when stacking a large number of spectra to search for a faint signal, it is important to test whether there are low-level correlations between the spectra, arising from systematic non-Gaussian effects (e.g. deconvolution errors from continuum sources, unmodelled changes in the antenna bandpass shapes over time, etc). If no correlations are present between the spectra, the RMS noise of the stacked spectrum is expected to decrease $\propto 1/\sqrt{N}$, where $N$ is the number of spectra that are stacked together. The presence of any correlations between the spectra would cause the RMS noise of the stacked spectrum to decline more slowly than $\propto 1/\sqrt{N}$.  We tested for such low-level correlations between the spectra by stacking smaller sub-samples of galaxies, randomly drawn from the full sample of 7,653 galaxies, and determining the dependence of the RMS noise on the number of stacked spectra. Specifically, we stacked random sub-samples containing 100, 200, 400, 800, 1,600, 3,200, and 6,400 galaxies and estimated the RMS noise on the \hi\ mass, for an assumed channel width of 270~\kmps. For each sub-sample, the RMS noise is computed in the same way as for the main stacked spectrum, i.e. by making 10,000 realizations of the stacked spectrum, using bootstrap re-sampling (with replacement) of the $N$ individual \hii\ spectra. The results are shown in Extended~Data~Figure~3, which plots the RMS noise on the stacked \hii\ spectrum against the number of stacked spectra. It is clear from the figure that the RMS noise on the stacked spectrum indeed decreases $\propto 1/\sqrt{N}$, indicating that there is no evidence for the presence of correlations between the \hii\ spectra. 

\section{Red galaxies and AGNs} 
We also stacked the \hii\ spectra from the sample of red DEEP2 galaxies, which were excluded from our main stack, to estimate their \hi \ mass. As noted earlier, the DEEP2 selection criteria preferentially picks out blue galaxies. There are only 1,469 red DEEP2 galaxies with reliable redshifts that lie within the spatial and spectral coverage of our uGMRT observations. After excluding AGNs and applying the same quality controls (described earlier) to the \hii \ spectra, we obtain a sample of 1,053 red galaxies. We stacked the \hii\ spectra of these 1,053 galaxies, following the approach described earlier, and find no evidence for a  detection of \hii\ emission. This implies the $3\sigma$ upper limit of $1.8 \times 10^{10} \ \Msun$ on the average \hi \ mass of red galaxies at $\langle z\rangle=0.95$. We note that this upper limit is larger than our estimate of the average \hi \ mass of blue star-forming galaxies, $\langle\MHI \rangle = (1.19 \pm 0.26) \times  10^{10} \ \Msun$.

We also combined the 1,053 red galaxies with the 7,653 blue galaxies to measure the average \hi \ mass of {\it all} galaxies in our sample. Stacking the \hii \ spectra of these 8,726 galaxies yields an \hi\ mass of $\langle\MHI \rangle = (0.97 \pm 0.24) \times  10^{10} \ \Msun$, consistent (within $1\sigma$ significance) with our result for the blue star-forming galaxies alone. The relatively small number of red galaxies due to the DEEP2 selection criteria implies that their inclusion in the stacking process does not appreciably affect our results. 

Finally, we examined the effect of including the 435 radio-bright AGNs on our estimate of the average \hi\ mass. After again applying the above quality controls to the \hii\ spectra of the 435 AGNs (yielding 368 usable spectra), we stacked the \hii \ spectra of the 7,653 blue galaxies and the 368 AGNs, obtaining an average \hi \ mass, $\langle\MHI \rangle = (1.02 \pm 0.26) \times  10^{10} \ \Msun$. This is again consistent, within statistical uncertainties, with our measurement of the average \hi\ mass of the blue galaxies alone. Again, the small number of AGNs in the DEEP2 sample implies that their retention in the sample does not substantially affect our results.

\section{The Effect of Source Confusion} 
For low angular resolution, the average \hii\ signal in a stacking experiment can include, in addition to the \hii\ emission from the target galaxies, \hii\ emission from gas in {\it companion} galaxies, lying within the synthesized beam and emitting at the same velocities as the target galaxy. Such ``source confusion'' can result in an over-estimation of the average \hi\ mass of the target galaxies. Simulations of \hii\ stacking experiments at $z \approx 0.7-0.758$ have found that source confusion does not dominate the signal from the target galaxies even with a resolution of $18''$ (corresponding to a physical size of $\approx 130$~kpc), with only $\approx 31\%$ of the stacked \hii\ signal arising from companion galaxies\cite{Elson16}. Our spatial resolution of 60~kpc is substantially smaller than this, and the effect of source confusion will thus be much lower. 

We used the S$^{3}$-SAX-Sky simulations to estimate the contamination in our detected \hii\ signal due to source confusion. The S$^{3}$-SAX-Sky simulation is based on semi-analytical models of galaxy evolution and provides a catalog of galaxies (including the \hi\ mass) out to $z\approx20$\cite{Obreschkow09}.  We retrieved the galaxies from the simulated catalog over a 1.2~square degree region and at redshifts $z = 0.74-1.45$, matched to the volume covered by our uGMRT observations; there are 657,421 galaxies in this volume.  The effect of source confusion is expected to be largest around massive galaxies, due to the strong clustering around these galaxies. Therefore, to get an upper limit on the effect of source confusion, we selected 7,653 galaxies with the largest \hi\ mass from the 657,421 simulated galaxies. The average \hi\ mass of these 7,653 simulated galaxies is $1.2\times10^{10} \ \Msun$, in excellent agreement with our estimate of the average \hi \ mass, $\langle\MHI \rangle = (1.19 \pm 0.26) \times  10^{10} \ \Msun$. Next, for each of these 7,653 simulated galaxies, we identified companion galaxies in the simulated catalog within the spatial and spectral resolution of our final stacked spectral cube, i.e. galaxies lying within 60~kpc and within 270 \ \kmps\ of the target galaxy. We assume that the \hii \ emission from the entire \hi \ mass of such companion galaxies will contribute to the stacked \hii\ signal. Even with these conservative assumptions (that would certainly over-estimate the contribution of source confusion to the measured \hi\ mass), we find that the companion galaxies contribute only $\approx2\%$ of the average \hi \ mass measurement. We thus conclude that the high spatial resolution (60~kpc) of the final spectral cube implies that our measurement of the average \hi\ mass of galaxies at $z=0.74-1.45$ is not appreciably affected by source confusion. \\ \\
\section{Determination of the SFR} 

We initially convolved the radio continuum images of the five uGMRT pointings to a uniform beam of 
FWHM=$5.5'' \times 5.5''$. We then followed a procedure similar to that discussed in the preceding 
section regarding the convolution of the \hii\ sub-cubes, in order to take into account any deviations
of the synthesized beam of each image from a Gaussian beam, so as to ensure that the convolved maps 
are in the correct units of Jy/beam. We hence convolved the point spread function of the continuum maps 
with the same kernel that was used in the images, and checked that the central pixel of the convolved 
point spread function is correctly normalized to unity. From these convolved images, we extracted 
$25'' \times 25''$ cutouts around the location of each of our 7,653 galaxies. For each galaxy, the 
flux density at each pixel was converted to its rest-frame 1.4~GHz luminosity density (${\rm L_{1.4 GHz}}$) 
at the galaxy's redshift, assuming a spectral index of $-0.8$\cite{Condon92}. These 1.4~GHz luminosity 
densities of the 7,653 galaxies were then stacked together, using a ``median stacking'' approach, computing 
the median 1.4~GHz luminosity density in each pixel from the galaxy sample. Such a median stacking procedure 
has been shown to be more robust to outliers (e.g. undetected AGNs in the sample) and deconvolution errors 
in the continuum images\cite{White07b}. Further, in cases (such as ours) of low signal-to-noise ratio ($\lesssim 1$) of the signal from individual objects of the sample, the median-stacking procedure yields the mean of the distribution\cite{White07}. We also applied the above procedure to locations offset by $100''$ from the true position of each galaxy, to search for possible systematic effects. The median-stacked 1.4~GHz luminosity density images, at the position of each DEEP2 galaxy and at the offset position, are shown in Extended~Data~Figure~4. The median-stacked 1.4~GHz luminosity density image at the position of the DEEP2 galaxies shows a clear detection of an unresolved continuum source at the centre of the image, with 1.4~GHz luminosity density ${\rm L_{1.4 GHz}}=(2.09\pm0.07) \times 10^{22}$~W~Hz$^{-1}$. No evidence is seen for systematic patterns in the offset stack. We converted our measured rest-frame 1.4~GHz luminosity density to an SFR estimate via a calibration derived from the radio-FIR correlation (but assuming a Chabrier IMF\cite{Yun01b}), SFR$(\Msun/\textrm{yr}) = (3.7 \pm 1.1) \times 10^{-22} \times {\rm L_{1.4 GHz} (W/Hz)}$. 
This yields a average SFR of $7.72 \pm 0.27 \ \Msun/\textrm{yr}$ for the 7,653 galaxies of our sample. 


\section{Stellar Masses for the DEEP2 Galaxies}
The stellar masses of the DEEP2 galaxies of our sample were inferred from a relation between the 
$\rm (U-B)$ color and the ratio of the stellar mass to the B-band luminosity, calibrated at $z\approx1$ using stellar masses estimated from K-band observations of a subset of the DEEP2 sample\cite{Weiner09b}. The RMS scatter of individual galaxies around the above relation is $\approx 0.3$~dex\cite{Weiner09b}.

\section{Reference sample at $z\approx0$}
A fair comparison of our results to those from the local Universe requires a 
uniformly-selected sample of nearby galaxies. The xGASS sample is a stellar 
mass-selected sample of galaxies at $z \approx 0$ with stellar masses $\Ms \geq 10^9 \, \Msun$, 
and with deep \hii\ emission studies yielding either a detection of \hii\ 
emission or, for non-detections, an \hi\ mass fraction relative to the stellar mass of $< 0.1$\cite{Catinella18b}. To enable a fair comparison with our results on blue galaxies at $z \approx 1$, 
we restricted the comparision sample at $z\approx0$ to blue galaxies from the xGASS sample, using the colour threshold $NUV-r < 4$. Next, 
the stellar mass distribution of the xGASS galaxies is different from that of our sample. We corrected 
for this effect by using the stellar mass distribution of our sample to determine weights when computing
the average stellar mass $\langle \Ms\rangle$, the average \hi\ mass $\langle\MHI\rangle$, and the 
average SFR $\langle {\rm SFR}\rangle$ for the xGASS galaxies. 

Finally, we note that $\approx4\%$ of the galaxies in our sample have $\Ms<10^9 \, \Msun$, a stellar mass range not covered by the xGASS survey; removing these galaxies from our sample has no substantial effect on the results presented in this work. 
\\
\section{Determination of $\OHI$}

The cosmological \hi\ mass density in galaxies at a redshift $z$ is defined as $\OHI(z) =\rho_{\rm HI}(z)/\rho_{\rm crit,0}$, where $\rho_{\rm HI}(z)$ is the comoving \hi\ mass density in galaxies at this redshift, and $\rho_{\rm crit,0}$ is the critical density of the Universe at $z=0$\cite{Wolfe05}. The estimation of $\rho_{\rm HI}(z)$ requires the measurement of the \hi\ mass of all galaxies in a given comoving volume~$V$ at the redshift of interest. In the nearby Universe, where \hii\ emission can be detected from individual galaxies, the cosmological \hi\ mass density is derived using $\rho_{\rm HI}(z)=\int_{-\infty}^{\infty} \Phi(\MHI) \,{\rm d}{\MHI}$ where $\Phi(\MHI)$, the ``\hi\ mass function", is the number density of galaxies per unit $\MHI$ at a given $\MHI$\cite{Jones18}.  However, in \hii\ stacking studies, we only have an estimate of the average \hi\ mass estimate for galaxies with spectroscopic redshifts; further, these galaxies are typically the brighter members of the population. In such experiments, the cosmological \hi\ mass density is usually computed using $\rho_{\rm HI}(z)=\int_{-\infty}^{\infty} \MHI({\rm M_X}).\phi({\rm M_X}) \,{\rm d}{\rm M_X}$, where $\MHI({\rm M_X})$ is the \hi\ mass of a galaxy at a given absolute magnitude ${\rm M_X}$ in the optical $\rm X$-band ($\rm X \equiv B, V, R, ...$), and $\phi({\rm M_X})$, the ``luminosity function", is the number density of galaxies per unit absolute magnitude ${\rm M_X}$ at a given ${\rm M_X}$. The luminosity function, $\phi({\rm M_X})$, is usually known from optical redshift surveys. The dependence of $\MHI$ on ${\rm M_X}$ is either (a)~characterized directly from the \hii\ stacking experiment by dividing the sample of galaxies in multiple sub-samples in ${\rm M_X}$ and finding the average \hi\ mass of galaxies in each of these sub-samples\cite{Hu19} , or (b)~assumed to be a power law, where only the normalization is constrained by the average \hi\ mass measured in the experiment\cite{Bera19}. We measure $\OHI$ in blue galaxies at $z\approx1$ by using a combination of these two approaches.

The DEEP2 galaxy sample is statistically unbiased up to a rest-frame B-band magnitude, $\rm M_B \leq -20$ at $z\approx1$\cite{Newman13}. We used this absolute magnitude-limited sample to estimate $\OHI$. There are 6,620 galaxies with $\rm M_B \leq -20$ at a mean redshift of $\langle z \rangle=1.06$ in our main sample of 7,653 blue star-forming galaxies.

The computation of \hi\ mass density, $\rho_{\rm HI}=\int_{-\infty}^{\infty} \MHI(\MB).\phi(\MB) \,{\rm d}\MB$, requires a knowledge of the dependence of $\MHI$ on the B-band magnitude, $\MB$, of the galaxies at $z\approx1$. In order to characterize the dependence of $\MHI$ on $\MB$, we split our sample of 6,620 galaxies with $\rm M_B \leq -20$ into two sub-samples separated by the median value of the distribution, $\MB=-21.042$, and stacked the \hii\ emission from the galaxies in each sub-sample to estimate the dependence of $\MHI$ on $\MB$. We find that the sub-sample of fainter galaxies, $\MB\ge-21.042$, has an average \hi\ mass of $\langle\MHI\rangle=(5.38 \pm 3.75) \times 10^9 \; \Msun$, while the sub-sample of brighter galaxies, $\MB\le-21.042$, has an average \hi\ mass of $\langle\MHI\rangle=(18.02 \pm 4.39) \times 10^9 \; \Msun$. Studies in the local Universe have found a relation between $\MHI$ and $\MB$ of the form

\begin{equation}
    \log \left[\MHI (\MB)\right] = {\rm K} - \beta \ \MB
\end{equation}
where ${\rm K}=2.89\pm0.11$ and $\beta=0.34\pm0.01$ at $z\approx0$\cite{Denes2014}. Assuming the same value of the slope, $\beta=0.34$, at $z\approx1$, we use our measurements of $\langle\MHI\rangle$ of galaxies in the two sub-samples to find the normalization of the relation to be ${\rm K}=2.88 \pm 0.11$; this is consistent, within statistical uncertainties, with the value of ${\rm K}=2.89\pm0.11$ measured at $z\approx0$. Extended Data Figure~5 shows the relation between $\MHI$ and $\MB$ at $z \approx 0$ [equation~1, with ${\rm K}=2.89$ and $\beta=0.34$\cite{Denes2014}] overlaid on our measurements of the average \hi\ mass of galaxies in the two $\MB$ sub-samples at $z\approx1$. Our measurements of the average $\MHI$ in the two sub-samples are consistent with the $\MHI-\MB$ relation measured at $z\approx 0$.  In passing, we note that our observations find evidence that the ratio of average \hi\ mass to stellar mass in blue star-forming galaxies changes from $z\approx1$ to $z\approx 0$, whereas the relation between $\MHI$ and $\MB$ appears to not change over the same redshift range. This could arise because $\MB$ is not a direct tracer of the stellar mass in galaxies.

We used the Schechter function fit to the B-band luminosity function of blue galaxies, $\phi(\MB)$, obtained from the DEEP2 survey\cite{Willmer06} to estimate the number density of galaxies at a given $\MB$. The Schechter function fits are available for three independent redshift bins: $0.8< z < 1.00$, $1.00< z < 1.20$ and $1.20 < z <1.40$. These bins are well matched to the redshift coverage of our observations and we thus take the mean of the three B-band luminosity functions to estimate the mean number density of galaxies at a given $\MB$ at $z=0.8-1.4$. Combining this with the $\MHI-\MB$ relation of equation~1 with ${\rm K}=2.88 \pm 0.11$ and  $\beta=0.34$, we obtain $\rho_{\rm HI}=\int_{-\infty}^{-20} \MHI(\MB).\phi(\MB) \,{\rm d}\MB = (3.15 \pm 0.79) \times 10^7 \, \Msun \, {\rm cMpc}^{-3}$. This yields $\OHI_{,\rm Bright}=(2.31 \pm 0.58) \times 10^{-4}$ for blue, star-forming galaxies with $\MB\leq -20$ at $\langle z \rangle=1.06$.

The above estimate of $\OHI_{,\rm Bright}$ does not include contributions from galaxies fainter than $\MB=-20$. To include these contributions, we assume that the relation between $\MHI$ and $\MB$ for galaxies with $\MB\leq -20$ at $\langle z \rangle=1.06$ can be extrapolated to fainter galaxies, with $\MB > -20$. With this extrapolation and the average B-band luminosity function at $z=0.8-1.4$, we obtain $\OHI= (4.5 \pm 1.1) \times 10^{-4}$, including contributions from {\it all} blue galaxies at $\langle z \rangle=1.06$. \\ \\

\section{Data Availability.}
The raw data reported in this paper are available through the GMRT archive:\\ (https://naps.ncra.tifr.res.in/goa) with project code: 35\_087. The analysed data files have a large size and are available from the corresponding author on reasonable request. The data displayed in Figure 1 are publicly available at https://github.com/chowdhuryaditya as a FITS file.

\section{Code Availability.}
The custom code used to calibrate the GMRT data is publicly available at https://github.com/chowdhuryaditya .


\begin{thebibliography}{10}
\expandafter\ifx\csname url\endcsname\relax
  \def\url#1{\texttt{#1}}\fi
\expandafter\ifx\csname urlprefix\endcsname\relax\def\urlprefix{URL }\fi
\providecommand{\bibinfo}[2]{#2}
\providecommand{\eprint}[2][]{\url{#2}}


\bibitem{Madau14}
\bibinfo{author}{{Madau}, P.} \& \bibinfo{author}{{Dickinson}, M.}
\newblock \bibinfo{title}{{Cosmic Star-Formation History}}.
\newblock \emph{\bibinfo{journal}{\araa}} \textbf{\bibinfo{volume}{52}},
  \bibinfo{pages}{415--486} (\bibinfo{year}{2014}).


\bibitem{Chengalur01}
\bibinfo{author}{{Chengalur}, J.~N.}, \bibinfo{author}{{Braun}, R.} \&
  \bibinfo{author}{{Wieringa}, M.}
\newblock \bibinfo{title}{{HI in Abell 3128}}.
\newblock \emph{\bibinfo{journal}{\aap}} \textbf{\bibinfo{volume}{372}},
  \bibinfo{pages}{768--774} (\bibinfo{year}{2001}).

\bibitem{Swarup91}
\bibinfo{author}{{Swarup}, G.} \emph{et~al.}
\newblock \bibinfo{title}{{The Giant Metre-Wave Radio Telescope}}.
\newblock \emph{\bibinfo{journal}{Current Science}}
\textbf{\bibinfo{volume}{60}}, \bibinfo{pages}{95--105} (\bibinfo{year}{1991}).

\bibitem{Gupta17}
\bibinfo{author}{{Gupta}, Y.} \emph{et~al.}
\newblock \bibinfo{title}{{The upgraded GMRT: opening new windows on the radio
  Universe}}.
\newblock \emph{\bibinfo{journal}{Current Science}}
  \textbf{\bibinfo{volume}{113}}, \bibinfo{pages}{707--714}
  (\bibinfo{year}{2017}).

\bibitem{Fernandez16}
\bibinfo{author}{{Fern{\'a}ndez}, X.} \emph{et~al.}
\newblock \bibinfo{title}{{Highest Redshift Image of Neutral Hydrogen in
  Emission: A CHILES Detection of a Starbursting Galaxy at z = 0.376}}.
\newblock \emph{\bibinfo{journal}{\apjl}} \textbf{\bibinfo{volume}{824}},
  \bibinfo{pages}{L1} (\bibinfo{year}{2016}).

\bibitem{Lah07}
\bibinfo{author}{{Lah}, P.} \emph{et~al.}
\newblock \bibinfo{title}{{The HI content of star-forming galaxies at z =  0.24}}.
\newblock \emph{\bibinfo{journal}{\mnras}} \textbf{\bibinfo{volume}{376}},
\bibinfo{pages}{1357--1366} (\bibinfo{year}{2007}).

\bibitem{Rhee13}
\bibinfo{author}{{Rhee}, J.} \emph{et~al.}
\newblock \bibinfo{title}{{Neutral atomic hydrogen (H I) gas evolution in field galaxies at z {$\sim$} 0.1 and {$\sim$}0.2}}.
\newblock \emph{\bibinfo{journal}{\mnras}} \textbf{\bibinfo{volume}{435}},
\bibinfo{pages}{2693--2706} (\bibinfo{year}{2013}).

\bibitem{Kanekar16}
\bibinfo{author}{{Kanekar}, N.}, \bibinfo{author}{{Sethi}, S.} \&
  \bibinfo{author}{{Dwarakanath}, K.~S.}
\newblock \bibinfo{title}{{The Gas Mass of Star-forming Galaxies at $z {\approx} 1.3$}}.
\newblock \emph{\bibinfo{journal}{\apjl}} \textbf{\bibinfo{volume}{818}},
  \bibinfo{pages}{L28} (\bibinfo{year}{2016}).

\bibitem{Bera19}
\bibinfo{author}{{Bera}, A.}, \bibinfo{author}{{Kanekar}, N.},
  \bibinfo{author}{{Chengalur}, J.~N.} \& \bibinfo{author}{{Bagla}, J.~S.}
\newblock \bibinfo{title}{{Atomic Hydrogen in Star-forming Galaxies at
  Intermediate Redshifts}}.
\newblock \emph{\bibinfo{journal}{\apjl}} \textbf{\bibinfo{volume}{882}},
  \bibinfo{pages}{L7} (\bibinfo{year}{2019}).

\bibitem{Newman13}
\bibinfo{author}{{Newman}, J.~A.} \emph{et~al.}
\newblock \bibinfo{title}{{The DEEP2 Galaxy Redshift Survey: Design,
  Observations, Data Reduction, and Redshifts}}.
\newblock \emph{\bibinfo{journal}{\apjs}} \textbf{\bibinfo{volume}{208}},
  \bibinfo{pages}{5} (\bibinfo{year}{2013}).

\bibitem{Weiner09}
\bibinfo{author}{{Weiner}, B.~J.} \emph{et~al.}
\newblock \bibinfo{title}{{Ubiquitous Outflows in DEEP2 Spectra of Star-Forming
  Galaxies at z = 1.4}}.
\newblock \emph{\bibinfo{journal}{\apj}} \textbf{\bibinfo{volume}{692}},
  \bibinfo{pages}{187--211} (\bibinfo{year}{2009}).

\bibitem{Catinella18}
\bibinfo{author}{{Catinella}, B.} \emph{et~al.}
\newblock \bibinfo{title}{{xGASS: total cold gas scaling relations and  molecular-to-atomic gas ratios of galaxies in the local Universe}}.
\newblock \emph{\bibinfo{journal}{\mnras}} \textbf{\bibinfo{volume}{476}},
\bibinfo{pages}{875--895} (\bibinfo{year}{2018}).


\bibitem{Yun01}
\bibinfo{author}{{Yun}, M.~S.}, \bibinfo{author}{{Reddy}, N.~A.} \&
  \bibinfo{author}{{Condon}, J.~J.}
\newblock \bibinfo{title}{{Radio Properties of Infrared-selected Galaxies in
  the IRAS 2 Jy Sample}}.
\newblock \emph{\bibinfo{journal}{\apj}} \textbf{\bibinfo{volume}{554}},
  \bibinfo{pages}{803--822} (\bibinfo{year}{2001}).

\bibitem{Brinchmann04}
\bibinfo{author}{{Brinchmann}, J.}, \bibinfo{author}{{Charlot}, S.},
\bibinfo{author}{{White}, S.~D.~M.}, \bibinfo{author}{{Tremonti}, C.},
\bibinfo{author}{{Kauffmann}, G.}, \bibinfo{author}{{Heckman}, T.} \&
\bibinfo{author}{{Brinkmann}, J.}
\newblock \bibinfo{title}{{The physical properties of star-forming galaxies in the low-redshift Universe}}.
\newblock \emph{\bibinfo{journal}{\mnras}} \textbf{\bibinfo{volume}{351}},
\bibinfo{pages}{1151--1179} (\bibinfo{year}{2004}).


\bibitem{Noeske07}
\bibinfo{author}{{Noeske}, K.~G.} \emph{et~al.}
\newblock \bibinfo{title}{{Star Formation in AEGIS Field Galaxies since z=1.1:  The Dominance of Gradually Declining Star Formation, and the Main Sequence of Star-forming Galaxies}}.
\newblock \emph{\bibinfo{journal}{\apjl}} \textbf{\bibinfo{volume}{660}},
\bibinfo{pages}{L43--L46} (\bibinfo{year}{2007}).
  
\bibitem{Rodighiero11}
\bibinfo{author}{{Rodighiero}, G.} \emph{et~al.}
\newblock \bibinfo{title}{{The Lesser Role of Starbursts in Star Formation at $z = 2$}}.
\newblock \emph{\bibinfo{journal}{\apjl}} \textbf{\bibinfo{volume}{739}},
\bibinfo{pages}{L40} (\bibinfo{year}{2011}).



\bibitem{White07}
\bibinfo{author}{{White}, R.~L.}, \bibinfo{author}{{Helfand}, D.~J.},
  \bibinfo{author}{{Becker}, R.~H.}, \bibinfo{author}{{Glikman}, E.} \&
  \bibinfo{author}{{de Vries}, W.}
\newblock \bibinfo{title}{{Signals from the Noise: Image Stacking for Quasars
  in the FIRST Survey}}.
\newblock \emph{\bibinfo{journal}{\apj}} \textbf{\bibinfo{volume}{654}},
  \bibinfo{pages}{99--114} (\bibinfo{year}{2007}).

\bibitem{Bera18}
\bibinfo{author}{{Bera}, A.}, \bibinfo{author}{{Kanekar}, N.},
  \bibinfo{author}{{Weiner}, B.~J.}, \bibinfo{author}{{Sethi}, S.} \&
  \bibinfo{author}{{Dwarakanath}, K.~S.}
\newblock \bibinfo{title}{{Probing Star Formation in Galaxies at z
  {\ensuremath{\approx}} 1 via a Giant Metrewave Radio Telescope Stacking
  Analysis}}.
\newblock \emph{\bibinfo{journal}{\apj}} \textbf{\bibinfo{volume}{865}},
  \bibinfo{pages}{39} (\bibinfo{year}{2018}).

\bibitem{Tacconi13}
\bibinfo{author}{{Tacconi}, L.~J.} \emph{et~al.}
\newblock \bibinfo{title}{{Phibss: Molecular Gas Content and Scaling Relations
  in z \~{} 1-3 Massive, Main-sequence Star-forming Galaxies}}.
\newblock \emph{\bibinfo{journal}{\apj}} \textbf{\bibinfo{volume}{768}},
  \bibinfo{pages}{74} (\bibinfo{year}{2013}).

\bibitem{Saintonge17}
\bibinfo{author}{{Saintonge}, A.} \emph{et~al.}
\newblock \bibinfo{title}{{xCOLD GASS: The Complete IRAM 30 m Legacy Survey of
  Molecular Gas for Galaxy Evolution Studies}}.
\newblock \emph{\bibinfo{journal}{\apjs}} \textbf{\bibinfo{volume}{233}},
  \bibinfo{pages}{22} (\bibinfo{year}{2017}).
  
\bibitem{Jones18}
\bibinfo{author}{{Jones}~M.~G.},
  \bibinfo{author}{{Haynes}~M.~P.}, 
  \bibinfo{author}{{Giovanelli}~R.},
  \bibinfo{author}{ {Moorman}~C.}
\newblock \bibinfo{title}{{The ALFALFA H I mass function: a dichotomy in the low-mass slope and a locally suppressed `knee' mass}}.
\newblock \emph{\bibinfo{journal}{\mnras}} \textbf{\bibinfo{volume}{477}},
  \bibinfo{pages}{2--17} (\bibinfo{year}{2018})

\bibitem{Wolfe05}
\bibinfo{author}{{Wolfe}, A.~M.}, \bibinfo{author}{{Gawiser}, E.} \&
\bibinfo{author}{{Prochaska}, J.~X.}
\newblock \bibinfo{title}{{Damped Ly {$\alpha$} Systems}}.
\newblock \emph{\bibinfo{journal}{\araa}} \textbf{\bibinfo{volume}{43}},
\bibinfo{pages}{861--918} (\bibinfo{year}{2005}).


\bibitem{Noterdaeme12}
\bibinfo{author}{{Noterdaeme}~P.} \emph{et~al.}
\newblock \bibinfo{title}{{Column density distribution and cosmological mass density of neutral gas: Sloan Digital Sky Survey-III Data Release 9}}.
\newblock \emph{\bibinfo{journal}{\aap}} \textbf{\bibinfo{volume}{547}},
  \bibinfo{pages}{L1} (\bibinfo{year}{2012})

\bibitem{Chang10}
\bibinfo{author}{{Chang}~T.~C.},
  \bibinfo{author}{{Pen}~U.-L}, 
  \bibinfo{author}{ {Bandura}~K.}
\newblock \bibinfo{title}{{An intensity map of hydrogen 21-cm emission at redshift z$\sim$0.8
}}.
\newblock \emph{\bibinfo{journal}{\nat}} \textbf{\bibinfo{volume}{466}},
  \bibinfo{pages}{463--465} (\bibinfo{year}{2010})
  
 \bibitem{Rao17}
\bibinfo{author}{{Rao}~ S.~M.}, \bibinfo{author}{{Turnshek}~D.~A.}, \bibinfo{author}{ {Sardane}~G.~M.}, \& \bibinfo{author}{ {Monier}~E.~M. }
\newblock \bibinfo{title}{{The statistical properties of neutral gas at $z < 1.65$ from UV measurements of Damped Lyman Alpha systems.
}}
\newblock \emph{\bibinfo{journal}{\mnras}} \textbf{\bibinfo{volume}{471}},
  \bibinfo{pages}{3428--3442} (\bibinfo{year}{2017})
  
\bibitem{Crighton15}
\bibinfo{author}{{Crighton}~N.~H.~M.} \emph{et~al.}
\newblock \bibinfo{title}{{The neutral hydrogen cosmological mass density at $z = 5$.
}}
\newblock \emph{\bibinfo{journal}{\mnras}} \textbf{\bibinfo{volume}{452}},
  \bibinfo{pages}{217--234} (\bibinfo{year}{2015})



\end{thebibliography}

\begin{thebibliography}{10}
\expandafter\ifx\csname url\endcsname\relax
  \def\url#1{\texttt{#1}}\fi
\expandafter\ifx\csname urlprefix\endcsname\relax\def\urlprefix{URL }\fi
\providecommand{\bibinfo}[2]{#2}
\providecommand{\eprint}[2][]{\url{#2}}
\makeatletter
\addtocounter{\@listctr}{26}
\makeatother

\bibitem{Madau14b}
\bibinfo{author}{{Madau}, P.} \& \bibinfo{author}{{Dickinson}, M.}
\newblock \bibinfo{title}{{Cosmic Star-Formation History}}.
\newblock \emph{\bibinfo{journal}{\araa}} \textbf{\bibinfo{volume}{52}},
  \bibinfo{pages}{415--486} (\bibinfo{year}{2014}).

\bibitem{Newman13b}
\bibinfo{author}{{Newman}, J.~A.} \emph{et~al.}
\newblock \bibinfo{title}{{The DEEP2 Galaxy Redshift Survey: Design,
  Observations, Data Reduction, and Redshifts}}.
\newblock \emph{\bibinfo{journal}{\apjs}} \textbf{\bibinfo{volume}{208}},
  \bibinfo{pages}{5} (\bibinfo{year}{2013}).

\bibitem{McMullin07}
\bibinfo{author}{{McMullin}, J.~P.}, \bibinfo{author}{{Waters}, B.},
  \bibinfo{author}{{Schiebel}, D.}, \bibinfo{author}{{Young}, W.} \&
  \bibinfo{author}{{Golap}, K.}
\newblock \emph{\bibinfo{title}{{CASA Architecture and Applications}}}, vol.
  \bibinfo{volume}{376} of \emph{\bibinfo{series}{Astronomical Society of the
  Pacific Conference Series}}, \bibinfo{pages}{127--130} (\bibinfo{year}{2007}).

\bibitem{Offringa12}
\bibinfo{author}{{Offringa}, A.~R.}, \bibinfo{author}{{van de Gronde}, J.~J.}
  \& \bibinfo{author}{{Roerdink}, J.~B.~T.~M.}
\newblock \bibinfo{title}{{A morphological algorithm for improving
  radio-frequency interference detection}}.
\newblock \emph{\bibinfo{journal}{\aap}} \textbf{\bibinfo{volume}{539}},
  \bibinfo{pages}{A95} (\bibinfo{year}{2012}).

\bibitem{Cornwell08}
\bibinfo{author}{{Cornwell}, T.~J.}, \bibinfo{author}{{Golap}, K.} \&
  \bibinfo{author}{{Bhatnagar}, S.}
\newblock \bibinfo{title}{{The Noncoplanar Baselines Effect in Radio
  Interferometry: The W-Projection Algorithm}}.
\newblock \emph{\bibinfo{journal}{IEEE Journal of Selected Topics in Signal
  Processing}} \textbf{\bibinfo{volume}{2}}, \bibinfo{pages}{647--657}
  (\bibinfo{year}{2008}).

\bibitem{Rau11}
\bibinfo{author}{{Rau}, U.} \& \bibinfo{author}{{Cornwell}, T.~J.}
\newblock \bibinfo{title}{{A multi-scale multi-frequency deconvolution
  algorithm for synthesis imaging in radio interferometry}}.
\newblock \emph{\bibinfo{journal}{\aap}} \textbf{\bibinfo{volume}{532}},
  \bibinfo{pages}{A71} (\bibinfo{year}{2011}).

\bibitem{Maddox13}
\bibinfo{author}{{Maddox}, N.}, \bibinfo{author}{{Hess}, K.~M.},
  \bibinfo{author}{{Blyth}, S.~L.} \& \bibinfo{author}{{Jarvis}, M.~J.}
\newblock \bibinfo{title}{{Comparison of H I and optical redshifts of galaxies
  - the impact of redshift uncertainties on spectral line stacking}}.
\newblock \emph{\bibinfo{journal}{\mnras}} \textbf{\bibinfo{volume}{433}},
  \bibinfo{pages}{2613--2625} (\bibinfo{year}{2013}).

\bibitem{Elson19}
\bibinfo{author}{{Elson}, E.~C.}, \bibinfo{author}{{Baker}, A.~J.} \&
  \bibinfo{author}{{Blyth}, S.~L.}
\newblock \bibinfo{title}{{On the uncertainties of results derived from H I
  spectral line stacking experiments}}.
\newblock \emph{\bibinfo{journal}{\mnras}} \textbf{\bibinfo{volume}{486}},
  \bibinfo{pages}{4894--4903} (\bibinfo{year}{2019}).


\bibitem{Willmer06}
\bibinfo{author}{{Willmer}, C.~N.~A.} \emph{et~al.}
\newblock \bibinfo{title}{{The Deep Evolutionary Exploratory Probe 2 Galaxy
  Redshift Survey: The Galaxy Luminosity Function to
  z\raisebox{-0.5ex}\textasciitilde1}}.
\newblock \emph{\bibinfo{journal}{\apj}} \textbf{\bibinfo{volume}{647}},
  \bibinfo{pages}{853--873} (\bibinfo{year}{2006}).

\bibitem{Condon02}
\bibinfo{author}{{Condon}, J.~J.}, \bibinfo{author}{{Cotton}, W.~D.} \&
  \bibinfo{author}{{Broderick}, J.~J.}
\newblock \bibinfo{title}{{Radio Sources and Star Formation in the Local
  Universe}}.
\newblock \emph{\bibinfo{journal}{\aj}} \textbf{\bibinfo{volume}{124}},
  \bibinfo{pages}{675--689} (\bibinfo{year}{2002}).

\bibitem{Wang16}
\bibinfo{author}{{Wang}, J.} \emph{et~al.}
\newblock \bibinfo{title}{{New lessons from the H I size-mass relation of
  galaxies}}.
\newblock \emph{\bibinfo{journal}{\mnras}} \textbf{\bibinfo{volume}{460}},
  \bibinfo{pages}{2143--2151} (\bibinfo{year}{2016}).

\bibitem{Bera19b}
\bibinfo{author}{{Bera}, A.}, \bibinfo{author}{{Kanekar}, N.},
  \bibinfo{author}{{Chengalur}, J.~N.} \& \bibinfo{author}{{Bagla}, J.~S.}
\newblock \bibinfo{title}{{Atomic Hydrogen in Star-forming Galaxies at
  Intermediate Redshifts}}.
\newblock \emph{\bibinfo{journal}{\apjl}} \textbf{\bibinfo{volume}{882}},
  \bibinfo{pages}{L7} (\bibinfo{year}{2019}).

\bibitem{Elson16}
\bibinfo{author}{{Elson}, E.~C.}, \bibinfo{author}{{Blyth}, S.~L.} \&
  \bibinfo{author}{{Baker}, A.~J.}
\newblock \bibinfo{title}{{Synthetic data products for future H I galaxy
  surveys: a tool for characterizing source confusion in spectral line stacking
  experiments}}.
\newblock \emph{\bibinfo{journal}{\mnras}} \textbf{\bibinfo{volume}{460}},
  \bibinfo{pages}{4366--4381} (\bibinfo{year}{2016}).

\bibitem{Obreschkow09}
\bibinfo{author}{{Obreschkow}, D.}, \bibinfo{author}{{Kl{\"o}ckner}, H.~R.},
  \bibinfo{author}{{Heywood}, I.}, \bibinfo{author}{{Levrier}, F.} \&
  \bibinfo{author}{{Rawlings}, S.}
\newblock \bibinfo{title}{{A Virtual Sky with Extragalactic H I and CO Lines
  for the Square Kilometre Array and the Atacama Large Millimeter/Submillimeter
  Array}}.
\newblock \emph{\bibinfo{journal}{\apj}} \textbf{\bibinfo{volume}{703}},
  \bibinfo{pages}{1890--1903} (\bibinfo{year}{2009}).

\bibitem{Condon92}
\bibinfo{author}{{Condon}, J.~J.}
\newblock \bibinfo{title}{{Radio emission from normal galaxies.}}
\newblock \emph{\bibinfo{journal}{\araa}} \textbf{\bibinfo{volume}{30}},
  \bibinfo{pages}{575--611} (\bibinfo{year}{1992}).

\bibitem{White07b}
\bibinfo{author}{{White}, R.~L.}, \bibinfo{author}{{Helfand}, D.~J.},
  \bibinfo{author}{{Becker}, R.~H.}, \bibinfo{author}{{Glikman}, E.} \&
  \bibinfo{author}{{de Vries}, W.}
\newblock \bibinfo{title}{{Signals from the Noise: Image Stacking for Quasars
  in the FIRST Survey}}.
\newblock \emph{\bibinfo{journal}{\apj}} \textbf{\bibinfo{volume}{654}},
  \bibinfo{pages}{99--114} (\bibinfo{year}{2007}).

\bibitem{Yun01b}
\bibinfo{author}{{Yun}, M.~S.}, \bibinfo{author}{{Reddy}, N.~A.} \&
  \bibinfo{author}{{Condon}, J.~J.}
\newblock \bibinfo{title}{{Radio Properties of Infrared-selected Galaxies in
  the IRAS 2 Jy Sample}}.
\newblock \emph{\bibinfo{journal}{\apj}} \textbf{\bibinfo{volume}{554}},
  \bibinfo{pages}{803--822} (\bibinfo{year}{2001}).

  
\bibitem{Weiner09b}
\bibinfo{author}{{Weiner}, B.~J.} \emph{et~al.}
\newblock \bibinfo{title}{{Ubiquitous Outflows in DEEP2 Spectra of Star-Forming
  Galaxies at z = 1.4}}.
\newblock \emph{\bibinfo{journal}{\apj}} \textbf{\bibinfo{volume}{692}},
  \bibinfo{pages}{187--211} (\bibinfo{year}{2009}).

\bibitem{Catinella18b}
\bibinfo{author}{{Catinella}, B.} \emph{et~al.}
\newblock \bibinfo{title}{{xGASS: total cold gas scaling relations and
  molecular-to-atomic gas ratios of galaxies in the local Universe}}.
\newblock \emph{\bibinfo{journal}{\mnras}} \textbf{\bibinfo{volume}{476}},
  \bibinfo{pages}{875--895} (\bibinfo{year}{2018}).

\bibitem{Hu19}
\bibinfo{author}{{Hu}, W.} \emph{et~al.}
\newblock \bibinfo{title}{{An accurate low-redshift measurement of the cosmic neutral hydrogen density}}.
\newblock \emph{\bibinfo{journal}{\mnras}} \textbf{\bibinfo{volume}{489}},
  \bibinfo{pages}{1619---1632} (\bibinfo{year}{2019}).


\bibitem{Denes2014}
\bibinfo{author}{{D{\'e}nes}, H.},
\bibinfo{author}{{Kilborn}, V.~A.},
\bibinfo{author}{{Koribalski}, B.~S.}
\newblock \bibinfo{title}{{New H I scaling relations to probe the H I content of galaxies via global H I-deficiency maps}}.
\newblock \emph{\bibinfo{journal}{\mnras}} \textbf{\bibinfo{volume}{444}},
  \bibinfo{pages}{667--681} (\bibinfo{year}{2014}).
  
  
\end{thebibliography}

\end{methods}

\begin{addendum}
 \item We thank the staff of the GMRT who have made these observations possible. 
The GMRT is run by the National Centre for Radio Astrophysics of the Tata Institute 
of Fundamental Research. N.K. acknowledges support from the Department of Science and 
Technology via a Swarnajayanti Fellowship (DST/SJF/PSA-01/2012-13). A.C., N.K. and J.N.C.
also acknowledge support from the Department of Atomic Energy, under project 12-R\&D-TFR-5.02-0700.
\item[Author Contributions] N.K. and A.C. wrote the GMRT proposal. A.C. carried out the analysis of the GMRT data, with N.K. and J.N.C. contributing to the data analysis. A.C., N.K., and J.N.C. contributed to the interpretation of the results. A.C. and N.K. wrote the manuscript. J.N.C., S.S., and K.S.D. contributed to the writing and the editing of the proposal and the manuscript.
 \item[Competing Interests] The authors declare that they have no competing financial interests.
 \item[Correspondence] Correspondence should be addressed to Nissim Kanekar.~(email: nkanekar@ncra.tifr.res.in).

\end{addendum}

\begin{center}
\includegraphics[width=\linewidth]{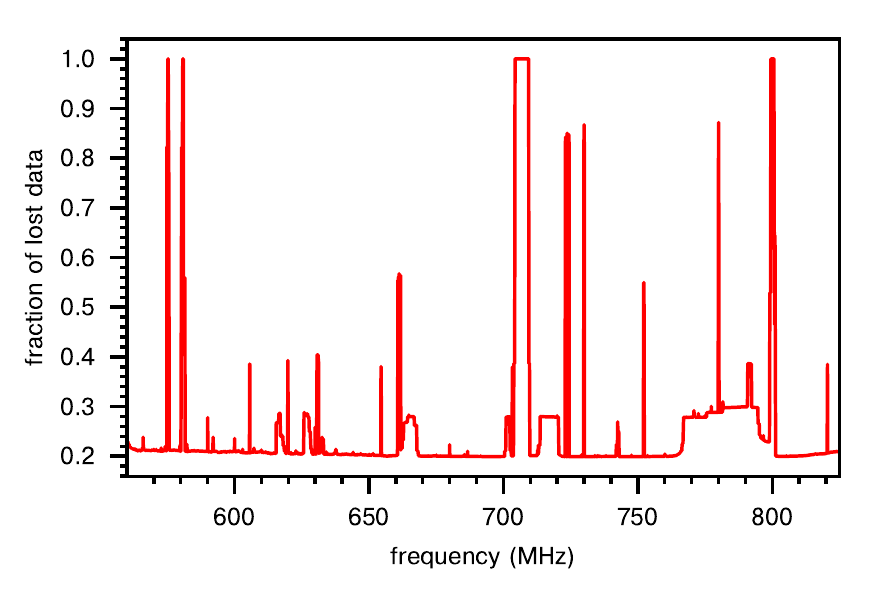}
\end{center}
\noindent{{\bf Extended Data Figure 1: The fraction of data excised across the observing band.} This includes all data lost due to time-variable effects, including RFI, malfunctioning antennas, power failures, etc. The plotted fraction of lost data was obtained by averaging over the $\approx 67$~hours of on-source time on the five DEEP2 sub-fields.}



 
\begin{center}
\includegraphics[width=\linewidth]{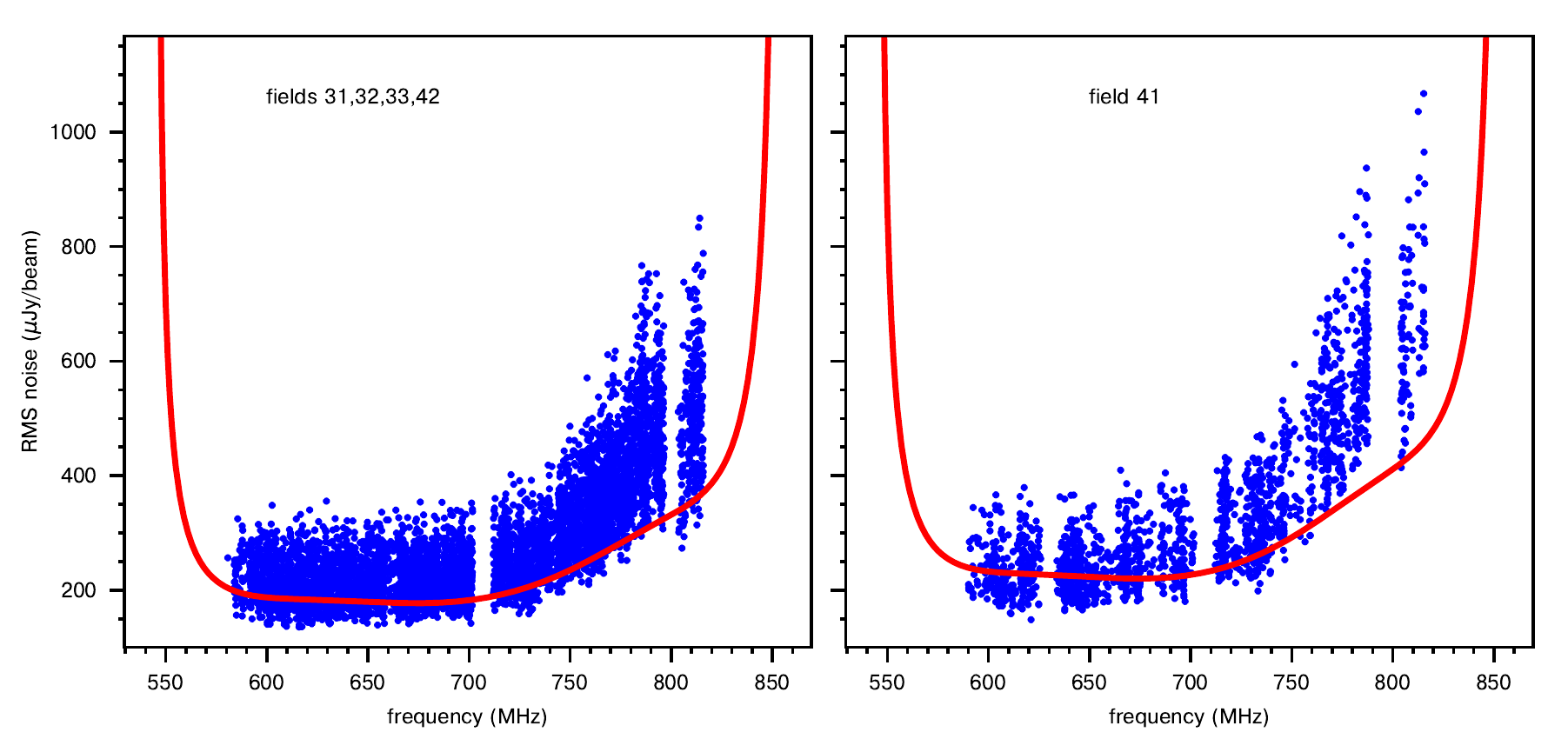}
\end{center}
\noindent{{\bf Extended Data Figure 2: Distribution of the spectral RMS noise for the 7,653 target galaxies.} The figure shows the RMS noise per 30~km/s channel on the \hii\ spectra of the 7,653 galaxies of the final sample. The left panel shows results for the galaxies in fields 31, 32, 33, $\&$ 42, each of which have $\approx 900$ minutes of on-source time. The right panel shows results for field~41, where the on-source time was $\approx 450$~minutes. The red curve in each panel shows the predicted RMS noise for the uGMRT array, after accounting for (1)~the on-source time, (2)~the fraction of data lost due to RFI and other effects, and (3)~the smoothing of the \hii\ cubes to the same spatial resolution at all redshifts, i.e. to coarser angular resolutions at higher frequencies. }

\begin{center}
\includegraphics[width=0.65\linewidth]{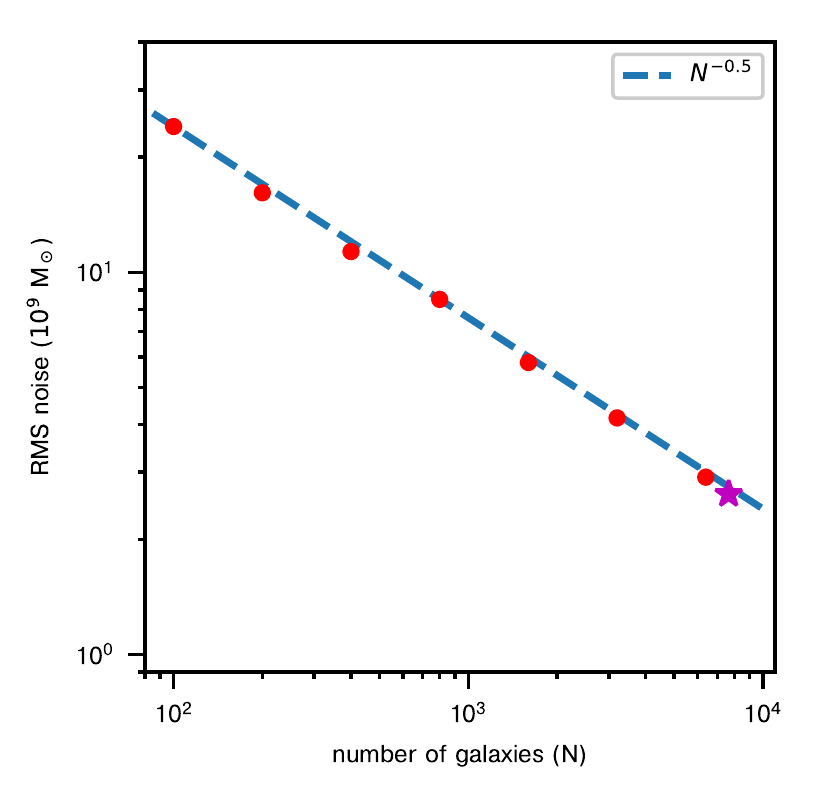}
\end{center}
\noindent{{\bf Extended Data Figure 3: Dependence of the stacked RMS noise on the number of galaxies.} The figure shows the RMS noise (in units of \hi\ mass sensitivity) on the stacked \hii\ spectrum as a function of the number of galaxies whose \hii\ spectra have been stacked together, assuming a velocity width of 270 \ \kmps. Each red circle shows the RMS noise from the spectrum of $N$ galaxies (with $N$=100, 200, 400, 800, 1,600, 3,200, and 6,400), randomly drawn from the full sample of 7,653 galaxies. The magenta star shows the RMS noise on the final stacked spectrum of 7,653 galaxies. The dashed blue line indicates the relation RMS=$N^{-0.5}$ (normalized to pass through the point with $N=100$), as expected if the 7,653 \hii\ spectra contain no correlations. The relation RMS~$\propto N^{-0.5}$ is an excellent match to the data points, implying that the \hii\ spectra show no evidence for the presence of systematic correlated non-Gaussian effects.}

\begin{center}
\includegraphics[width=\linewidth]{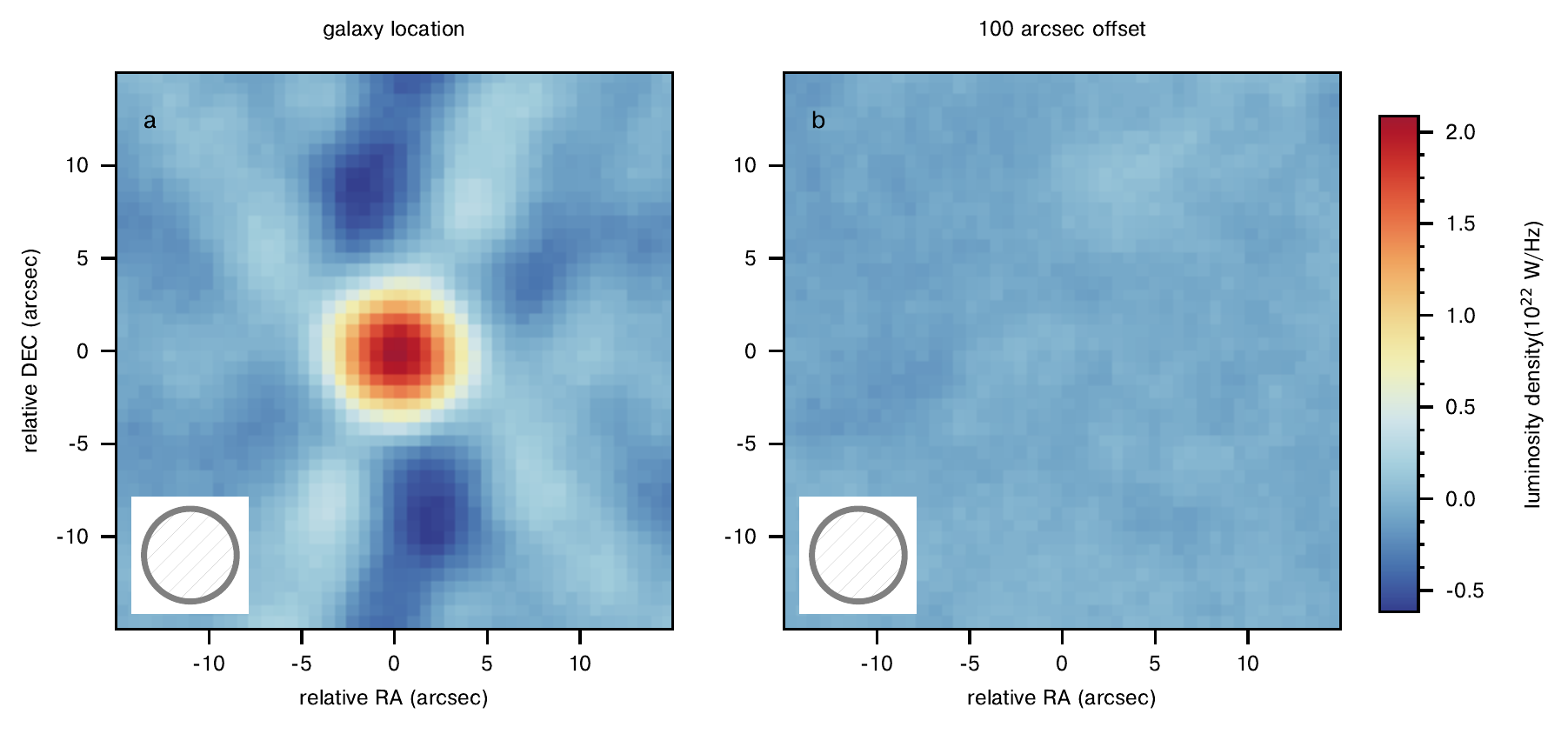}
\end{center}
\noindent{{\bf Extended Data Figure 4: The stacked 1.4~GHz continuum emission from our galaxies and offset positions.} The figure shows the average rest-frame 1.4~GHz luminosity density of the 7,653 main-sequence DEEP2 galaxies, obtained by median-stacking the 1.4~GHz radio continuum emission at the location of (a) each individual galaxy, and (b) at a location $100''$ offset from each galaxy. A clear ($29\sigma$ significance) detection is visible at the location of the DEEP2 galaxies, while the stack at offset positions shows no evidence for either emission or any systematic patterns. The circle in the bottom left corner represents the $5.5''$ beam to which all continuum images were convolved before the stacking. The patterns visible in the left panel around the central bright source arise from the effective point spread function of the stacked image.}

\begin{center}
\includegraphics[width=0.7\linewidth]{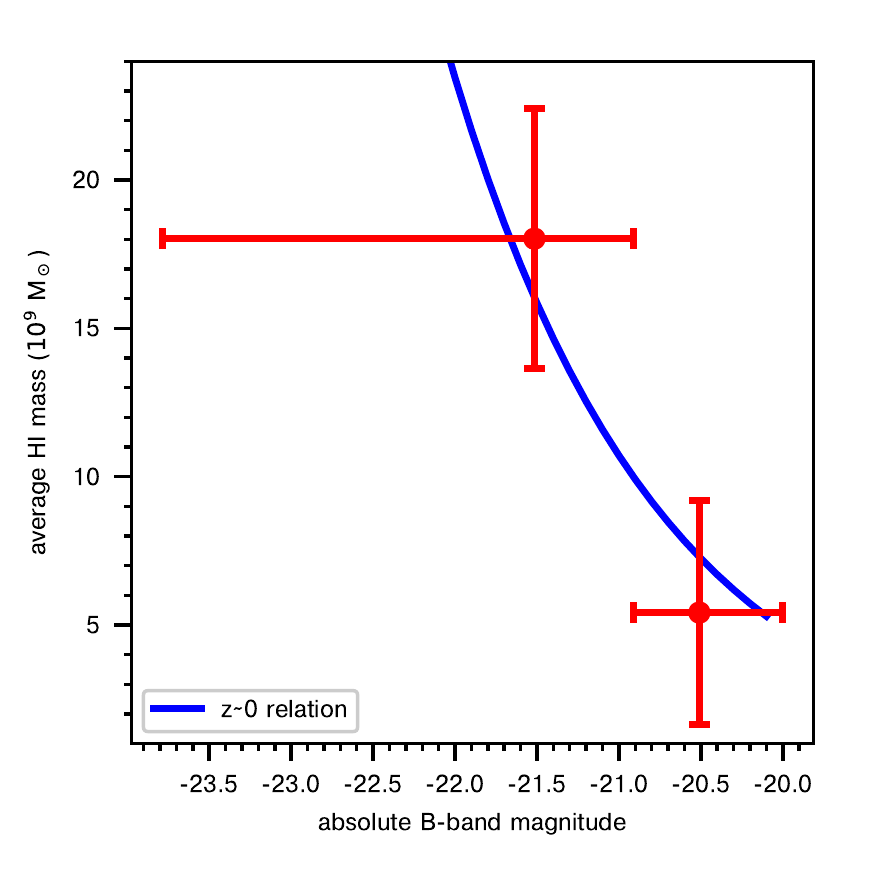}
\end{center}
\noindent {{\bf Extended Data Figure 5: The relation between \hi\ mass and absolute blue magnitude.} The figure shows the relation between average \hi\ mass and absolute B-band magnitude for galaxies with $\MB\le-20$ at $\langle z\rangle=1.06$. The red points shows the average \hi\ mass, obtained by stacking the \hii\ emission, of blue galaxies in two $\MB$ bins (separated at the median, $\MB = -21.042$) at $\langle z\rangle=1.06$. The solid blue curve shows the relation between $\MHI$ and $\MB$ in the local Universe\cite{Denes2014}. Our measurements at $\langle z\rangle=1.06$ are consistent with the $\MB-\MHI$ relationship at $z\approx0$. }

\begin{center}
\begin{tabular}{|c|c|c|c|c|c|c|}
\hline
\hline
   DEEP2  & Right Ascension& Declination& On-Source & Beam& $\sigma_\textrm{RMS}$ & Number of  \\
   sub-field  & (J2000) & (J2000) & Time & &  & Galaxies  \\ \hline\hline
   Field-31 & $23$h$26$m$52.8$s & $+00\degree08'25.7''$ & 861 min & $4.3''\times 4.0''$ & 5.7 $\mu$Jy/Bm & 1,353\\  
   Field-32 & $23$h$29$m$49.9$s & $+00\degree12'12.7''$ &872 min & $5.4''\times 4.5''$ & 5.2 $\mu$Jy/Bm & 1,310 \\ 
   Field-33 & $23$h$32$m$58.7$s & $+00\degree08'22.7''$ &919 min & $4.8''\times 4.2''$ & 5.7 $\mu$Jy/Bm & 1,283 \\ 
   Field-41 & $02$h$28$m$24.0$s & $+00\degree35'27.6''$ &450 min & $4.8''\times 4.7''$ & 8.0 $\mu$Jy/Bm & 1,872\\ 
   Field-42 & $02$h$30$m$48.0$s  & $+00\degree35'15.0''$&902 min & $5.0''\times 3.8''$ & 5.8 $\mu$Jy/Bm & 1,835\\ 
\hline
\hline
\end{tabular}
\end{center}
\noindent{{\bf Extended Data Table 1: Summary of the GMRT observations.} For each DEEP2 sub-field that was observed with 
the GMRT $550-850$~MHz receivers, the columns provide the J2000 coordinates of the GMRT pointing, 
the total on-source time (in minutes), the synthesized beam obtained in the continuum image, the RMS noise (in $\mu$Jy/Bm) measured on the continuum image away from detected sources ($\sigma_\textrm{RMS}$), and the number of galaxies from the sub-field that were included in the average $\MHI$ and SFR measurements.} 

\end{document}